\documentclass{article}
   \usepackage{graphicx}
   \usepackage{amssymb}
   \usepackage{epstopdf}
   \usepackage{apjgalley}
   \usepackage{multicol}
   \usepackage{psfig}
\font\fiverm=cmr5             \font\sevenrm=cmr7

          \font\sixrm=cmr6       
                     
\def\app{Astroparticle Phys.}                   
\def\ssr{Space Sci. Rev.}                       
\def\teq#1{$\, #1\,$}                           
\def\erg{\varepsilon}

\def\lambar{\lambda\llap {--}}
\def\fsc{\alpha_{\hbox{\sevenrm f}}}                                
\def\dover#1#2{\hbox{${{\displaystyle#1 \vphantom{(} }\over{
   \displaystyle #2 \vphantom{(} }}$}}                

\def\sigt{\hbox{$\sigma_{\hbox{\fiverm T}}$}}  
\def\taut{\hbox{$\tau_{\hbox{\fiverm T}}$}}

\def\eb{\erg_{\hbox{\fiverm B}}} 
\def\es{\erg_{\hbox{\fiverm S}}}

\def\gammat{\gamma_{\hbox{\fiverm T}}}   
\def\gammin{\gamma_{\hbox{\sevenrm min}}}   
\def\gammax{\gamma_{\hbox{\sevenrm max}}}   
\def\chit{\chi_{\hbox{\fiverm T}}}   
\def\thetamin{\theta_{\hbox{\sevenrm min}}}   
\def\thetamax{\theta_{\hbox{\sevenrm max}}}   
\def\ndotsyn{{\dot n}_{\hbox{\fiverm S}}}
\def\ndotpas{{\dot n}_{\hbox{\fiverm PAS}}}
\def\Ndotsyn{{\dot N}_{\hbox{\sevenrm syn}}}
\def\ndotIC{{\dot n}_{\hbox{\fiverm IC}}}

\def\ndotssc{{\dot n}_{\hbox{\fiverm SSC}}}
\def\Ndotssc{{\dot N}_{\hbox{\fiverm SSC}}}
\def\Ndotsscth{{\dot N}_{\hbox{\fiverm SSC,TH}}}
\def\Psyn{P_{\hbox{\fiverm S}}}
\def\us{u_{\hbox{\fiverm S}}} 
\def\ub{u_{\hbox{\fiverm B}}} 
\def\tesc{t_{\hbox{\sevenrm esc}}}      
\def\zmax{z_{\hbox{\sixrm max}}}
\def\today{\ifcase\month\or
  January\or February\or March\or April\or May\or June\or
  July\or August\or September\or October\or November\or
  December\fi
  \space\number\day, \number\year}
\begin{document}
%
%
\newcommand{\vol}[2]{$\,$\rm #1\rm , #2.}                 
\newcommand{\figureout}[5]{\centerline{}
   \centerline{\hskip #3in \psfig{figure=#1,width=#2in}}
   \vspace{#4in} \figcaption{#5} }
\newcommand{\twofigureout}[3]{\centerline{}
   \centerline{\psfig{figure=#1,width=3.4in}
        \hskip 0.5truein \psfig{figure=#2,width=3.4in}}
    \figcaption{#3} }    
\newcommand{\figureoutpdf}[5]{\centerline{}
   \centerline{\hspace{#3in} \includegraphics[width=#2truein]{#1}}
   \vspace{#4truein} \figcaption{#5} \centerline{} }
\newcommand{\twofigureoutpdf}[3]{\centerline{}
   \centerline{\includegraphics[width=3.4truein]{#1}
        \hspace{0.5truein} \includegraphics[width=3.4truein]{#2}}
        \vspace{-0.2truein}
    \figcaption{#3} }    

\newcommand{\figureoutfixed}[2]{\centerline{}
   \centerline{\psfig{figure=#1,width=5.5in}}
    \figcaption{#2}\clearpage }
\newcommand{\figureoutsmall}[2]{\centerline{\psfig{figure=#1,width=5.0in}}
    \figcaption{#2}\clearpage }
\newcommand{\figureoutvsmall}[2]{\centerline{\psfig{figure=#1,width=4.0in}}
    \figcaption{#2}\clearpage }
\newcommand{\tableout}[4]{\vskip 0.3truecm \centerline{\rm TABLE #1\rm}
   \vskip 0.2truecm\centerline{\rm #2\rm}   
   \vskip -0.3truecm  \begin{displaymath} #3 \end{displaymath} 
   \noindent \rm #4\rm\vskip 0.1truecm } 
%
%
%
%

\title{A STUDY OF PROMPT EMISSION MECHANISMS IN GAMMA-RAY BURSTS}

   \author{Matthew G. Baring}
   \affil{Department of Physics and Astronomy MS-108, \\
      Rice University, P.O. Box 1892, Houston, TX 77251, U.S.A.\\
      \it baring@rice.edu\rm}
    
    \and
    
    \author{Matthew L. Braby}
    \affil{Department of Physics, Washington University\\
      One Brookings Drive, St. Louis, MO 63130, U.S.A.\\
      \it mbraby@hbar.wustl.edu\rm}
\slugcomment{To appear in \it The Astrophysical Journal\rm , 
 Vol 613, September 20, 2004 issue.}
%

\begin{abstract}  
The principal paradigm for the generation of the non-thermal particles
that are responsible for the prompt emission of gamma-ray bursts invokes
diffusive shock acceleration at shocks internal to the dynamic
ultrarelativistic outflow.  This paper explores expectations for burst
emission spectra arising from shock acceleration theory in the limit of
particles cooling much slower than their acceleration.  Parametric fits
to burst spectra obtained by the Compton Gamma-Ray Observatory (CGRO)
are explored for the cases of the synchrotron, inverse Compton and
synchrotron self-Compton radiation mechanisms, using a linear
combination of thermal and non-thermal electron populations.  These fits
demand that the preponderance of electrons that are responsible for the
prompt emission reside in an intrinsically non-thermal population,
strongly contrasting particle distributions obtained from acceleration
simulations. This implies a potential conflict for acceleration
scenarios where the non-thermal electrons are drawn directly from a
thermal gas, unless radiative efficiencies only become significant at
highly superthermal energies. It is also found that the CGRO data
preclude effective spectroscopic discrimination between the synchrotron
and inverse Compton mechanisms.   This situation may be resolved with
future missions probing gamma-ray bursts, namely Swift and GLAST. 
However, the synchrotron self-Compton (SSC) spectrum is
characteristically too broad near the \teq{\nu F_{\nu}} peak to viably
account for bursts such as GRB 910601, GRB 910814 and GRB 990123. It is
concluded that the SSC mechanism may be generally incompatible with
differential burst spectra steeper than around \teq{E^{-2.5}} above the
peak, unless the synchrotron component is strongly self-absorbed.
\end{abstract}  
\section{INTRODUCTION}
\label{sec:intro}

Cosmological gamma-ray bursts are one of the most interesting and exotic
phenomena in astrophysics.  In standard burst models (e.g. see Piran
1999; M\'esz\'aros 2001 for reviews), the rapidly expanding fireball
decelerates, converting the internal energy of the hot plasma into
kinetic energy of the beamed, relativistically moving ejecta and
electron-positron pairs. At the point where the fireball becomes
optically thin and the gamma-ray burst (GRB) we see is emitted, the
matter will not emit non-thermal gamma-rays unless some mechanism can
efficiently re-convert the kinetic energy back into internal energy,
i.e., unless some particle acceleration process takes place.  Diffusive
shock acceleration is widely believed to be this mechanism (e.g., Rees
\& M\'esz\'aros 1992; Piran 1999) for the prompt emission; it is also
likely to be a key element for the underlying physics of X-ray and
optical flashes and burst afterglows.

An important question concerning burst models is whether or not their
prompt emission can be accurately described by a detailed shock
acceleration analysis, and if so, how is the pertinent parameter space
of models limited by the observations?  Tavani (1996a,b) proposed a
synchrotron shock emission model that invoked such diffusive
acceleration as the means for generating non-thermal distributions. This
work has been an interpretative driver in the field, due in part to the
impressive model fits of data he obtained. Yet, Tavani's model and many
subsequent works do not treat the critical involvement the acceleration
process has on shaping the relationship between the thermal and
non-thermal portions of the electron distribution.  The normal approach
is to simplify the distribution to a truncated or broken power-law,
occasionally with an additional thermal component. The viability of the
synchrotron model, or emission scenarios that invoke any other radiation
mechanism, hinges on such subtleties of the particle distribution, which
are inextricably linked to the nature of shock acceleration, diffusive
or coherent.  In addition, they must provide compatibility with the hard
X-ray spectral index, an issue raised by Preece et al. (1998; 2000), who
observed that around 1/3 of all BATSE bursts exhibited spectra that were
too flat below the \teq{\nu F_{\nu}} peak to accommodate a synchrotron
interpretation.

This paper explores the general expectations for burst emission spectra
from shock acceleration theory.  Parametric descriptions for a linear
combination of thermal and non-thermal distributions are used to obtain
fits to burst spectra for the cases of the synchrotron, inverse Compton
and synchrotron self-Compton (SSC) mechanisms, and these distributions
are interpreted in the context of known results from acceleration
simulations.  Cases of particles cooling much slower than their
acceleration are treated in this exposition; justification for these are
discussed in Section~\ref{sec:cooling}. The focus is on GRBs observed by
the Compton Gamma-Ray Observatory, specifically those with EGRET
detections that provide broad spectral coverage and therefore more
constrained fits around the spectral peak.  It is found that acceptable
fits are only possible with a marked dominance of non-thermal electrons,
contrasting the particle distributions obtained in acceleration
simulations.  This poses a general problem for any burst acceleration
paradigm where the non-thermal population is drawn probabilistically
from a thermal gas.

In addition, fits are found to be more or less equally acceptable for
the synchrotron and inverse Compton processes; it is anticipated that
Swift will provide a number of bright bursts where spectral fitting down
into the X-ray band might enable discrimination between such emission
mechanisms.  However, at this juncture, spectroscopic discrimination
against unabsorbed synchrotron self-Compton scenarios seems probable,
given that the characteristically broad SSC spectra in \teq{\nu F_{\nu}}
space are difficult to reconcile with CGRO data.  In service of this
analysis, a compact formalism for SSC emissivities from thermal and
truncated power-law electron distributions is present in the Appendix.
It is anticipated that GLAST's broad spectral coverage from \teq{\sim
10}keV to \teq{\sim 100}GeV, supplied by the Gamma-Ray Burst Monitor in
the BATSE band and the Large Area Telescope in and above the EGRET band,
will render GLAST a powerful tool in constraining and interpreting burst
spectral properties in the future.


\section{SYNCHROTRON EMISSION AND ITS INTERPRETATION}
 \label{sec:synchrotron}

This section explores the relationship of synchrotron spectra to 
the electron energy and angular distributions, inferring Lorentz
factor distributions using data from the bright burst GRB 910503 
as a representative case study.

\subsection{Synchrotron Formalism and Electron Distributions}
 \label{sec:formalism}

The formalism for synchrotron radiation in optically thin environs is
standard and readily available in many textbooks, including the
treatments of Bekefi (1966), Jackson (1975) and Rybicki \& Lightman
(1979).  Throughout this paper, the standard convention for the
labelling of the photon polarizations will be adopted, namely that
\teq{\parallel} refers to the state with the photon's {\it electric}
field vector parallel to the plane containing the magnetic field and the
photon's momentum vector, while \teq{\perp} denotes the photon's
electric field vector being normal to this plane.

The classical (photon) angle-integrated emissivities for the
\teq{\perp} and \teq{\parallel} polarization for monoenergetic
electrons of Lorentz factor \teq{\gamma} moving at angle \teq{\theta}
to the field are (e.g.  Westfold 1959; Jackson 1975; Rybicki and
Lightman 1979)
\begin{eqnarray}
 {\dot n}_{\perp}(\gamma ,\;\erg ) & = & {\dot N}_{\theta}
   \; \Biggl\{ \int_{\erg /\erg_c \sin\theta}^{\infty}
   K_{5/3}(x)\, dx +
   K_{2/3}\biggl(\dover{\erg}{\erg_c\sin\theta}\biggr)\Biggr\}\;\; ,
   \nonumber \\
   & & \\
 {\dot n}_{\parallel}(\gamma ,\;\erg ) & = & {\dot N}_{\theta}
   \; \Biggl\{ \int_{\erg /\erg_c \sin\theta}^{\infty}
   K_{5/3}(x)\, dx - 
   K_{2/3}\Bigl(\dover{\erg}{\erg_c\sin\theta}\Bigr)\Biggr\}\;\; ,
   \nonumber
  \label{eq:synchrate}
\end{eqnarray}
in units of \teq{sec^{-1}}, where the \teq{K_{\nu}} are modified Bessel
functions of the second kind.  The generally higher probability of emitting
\teq{\perp} photons than \teq{\parallel} ones reflects the intrinsic
dipolar radiative nature of accelerating/oscillating charges; the
polarization is elliptical for photon propagation along the field.  The
characteristic synchrotron rate factor that scales these emissivities is
\begin{equation}
   {\dot N}_{\theta}\; =\; \dover{\fsc}{2\pi\sqrt{3}}\,
   \dover{m_ec^2}{\hbar}\; \dover{\sin\theta}{\gamma^2}\quad .
 \label{eq:Ndotdef}
\end{equation}
Here, \teq{\fsc =e^2/(\hbar c)} is the fine structure constant.  The
photon energies \teq{\erg_c} that appear in these functions that form
the scaling for the energies of emission are [with the electron
Compton wavelength over \teq{2\pi} being \teq{\lambar =\hbar/(m_ec)}]:
\begin{equation}
   \erg_c\; =\;\dover{3}{2}\; \dover{B}{B_{\rm cr}}\,\gamma^2
   \; =\;\dover{3}{2}\,\gamma^3\, \dover{\lambar}{r_{\hbox{\fiverm L}}}
  \label{eq:ergcsynch}
\end{equation}
where \teq{r_{\hbox{\fiverm L}}} is the particle's Larmor radius, and
\teq{B_{\rm cr} = 4.41 \times 10^{13}}Gauss is the quantum critical
field.  Note that since \teq{\gamma\gg 1}, the photon angles to the
field form a narrow distribution about \teq{\theta}.  Furthermore, the
expressions in Eq.~(\ref{eq:synchrate}) are appropriate for cases where
the particle momenta are not closely aligned with the magnetic field;
situations of such near alignment are discussed in
Section~\ref{sec:pitch_synch} below.

The total synchrotron emissivity is obtained by integrating over the
energy and angular distribution of the particles.  For electron
distributions \teq{n_e(\gamma,\, \theta)} in units of \teq{cm^{-3}},
summing over polarizations forms the expression for the emissivity that
will be used mostly throughout this paper (in units of \teq{cm^{-3}\,
sec^{-1}}):
\begin{equation}
   \ndotsyn (\erg )\; =\; \int d\gamma \int\sin\theta\, d\theta\;\;
   n_e(\gamma,\, \theta)\;
   \sum_{p=\perp,\parallel} {\dot n}_p(\gamma ,\;\erg )\;\; .
  \label{eq:synch_emiss}
\end{equation}
This is the differential photon spectrum; frequently in this work, 
reference will be made to the so-called \teq{\nu F_\nu} spectrum,
a spectral energy distribution that is proportional to
\teq{\erg^2\ndotsyn (\erg )}.

Physical diagnostics on the GRB environment are obtained by comparing
the emission spectrum in Eq.~(\ref{eq:synch_emiss}) with observations.
The obvious database of choice comprises BATSE observations from the
Compton Gamma-Ray Observatory (CGRO), specifically the BATSE
spectroscopy catalog of Preece et al. (2000) with its pubicly-accessible
electronic compilation. Yet this data possesses a principal limitation,
namely that imposed by the restricted energy band of BATSE.  Inferences
of particle distributions are generally on a more secure footing when
using broader burst spectra, particularly in domains of significant
spectra structure, which is the intrinsic nature of the BATSE pool. 
Thus, the two decades of energy afforded by BATSE often are insufficient
to describe the nature of the non-thermal particle distribution invoked
in models.  Data from the EGRET experiment on CGRO suitably augment this
situation and extend the dynamic range by 2--4 orders of magnitude in
energy.  Accordingly, a focus on EGRET bursts is insightful and offers
more secure determinations of particle distributions.  Use is therefore
made of spectral compilations of BATSE, COMPTEL and EGRET data that are
offered in Schaefer et al. (1998).  Specifically, a case study is made
here of GRB 910503, observing that the inferences to be made generally
extend to the handful of other EGRET bursts, and also to a large number
of bright BATSE bursts that possess no confirmed EGRET detections.  

Note that the focus here is primarily on time-integrated spectra.  Given
the complexity and diversity of  burst time profiles, and associated
variations in spectral shape and the well-known hard-to-soft evolution,
detailed spectral fits will depend on the time window chosen.  A classic
example for this is provided by the gradual emergence of a hard
gamma-ray component in the spectrum of GRB 941017 (Gonzalez, et al.
2003).  The main conclusions of this paper apply to the mean properties
of the emitting particle distributions, and are striking enough to
indicate their applicability for most or nearly all of the duration of
the bursts studied herein.

A variety of particle distributions are possible in modeling burst
spectra.  Clearly, all should contain a non-thermal component, since the
data are clearly non-thermal:  purely isothermal models do not work
(Pacynski 1986; Goodman 1986), though constructed convolutions of
thermal spectra of different temperatures are possible.  The emphasis
here is motivated by physical interpretation, and the leading contender
for non-thermal particle energization in bursts is diffusive
acceleration at relativistic shocks in the burst outflow.  Such
acceleration is well studied in non-relativistic flows, and is widely
believed to originate with thermal particles that are diffusively
transported when interacting with field turbulence in the shock
environs.  Such transport effects many crossings of the shock layer
(e.g. see Drury 1983; Blandford \& Eichler 1987; Jones \& Ellison 1991
for reviews), that lead to a friction term in the momentum evolution,
i.e. acceleration.  While other acceleration models such as reconnection
and coherent electrodynamic energization exist and are appropriate in a
variety of astrophysical environments (e.g. the solar corona, pulsar
magnetospheres), diffusive shock acceleration is the most widely invoked
in astrophysical models; it forms the centerpiece of discussion here as
well as for many investigations of bursts.

Tavani (1996a,b) proposed a shock emission model that invoked such
diffusive acceleration as the means for generating non-thermal
distributions. This work has been an interpretative driver in the field,
in no small part due to the impressive model fits of data obtained.
The principal purpose of this Section is to delve deeper into the
physical implications of such data fitting.  Accordingly, it is
appropriate to use an electron distribution similar to Tavani's (1996a,b)
to add insight to the discussion.  Here, the quasi-isotropic
electron distribution
\begin{equation}
   n_e(\gamma,\, \theta)\; =\; n_{\theta} \Biggl\lbrack\;
    \biggl( \dover{\gamma}{\gammat} \biggr)^2\,
   e^{(-\gamma/\gammat )} + \epsilon \,
    \biggl( \dover{\gamma}{\gammat} \biggr)^{-\delta}\,
    \Theta \biggl( \dover{\gamma}{\eta\gammat} \biggr)\, \Biggr\rbrack\, ,
 \label{eq:elec_dist}
\end{equation} 
is adopted, where \teq{\Theta (x)} is a step function with \teq{\Theta
(x)=1} for \teq{x\geq 1} and zero otherwise, and \teq{\gammat
=kT/m_ec^2} is the system or post-shock temperature.  The factor
\teq{\epsilon} directly relates to the efficiency of acceleration, and
is a parameter in the fitting protocol, as are \teq{\gammat} (assumed
much greater then unity), the dimensionless \teq{\eta}, the non-thermal
index \teq{\delta}, and the normalization factor \teq{n_{\theta}}.  The
ensuing discussion will provide physical grounds justifying this choice
of distribution in shocked environs; other possibilities with different
mandates include the broken power-law adopted by Lloyd \& Petrosian
(2000).  Note that \teq{n_{\theta}} is independent of \teq{\theta} for
true isotropy.  In many cases in this paper, \teq{n_{\theta}\propto
\delta (\theta -\pi/2)} will be chosen for simplicity, since this
corresponds to the dominant radiative contribution from the synchrotron
mechanism.

The parameter \teq{\eta} is introduced to generalize from Tavani's
(1996a,b) fits, permitting the non-thermal distribution to emerge perhaps
at truly super-thermal energies, rather than exactly in the thermal
peak (Tavani's restricted case).  Typical acceleration spectra in
non-relativistic shocks assume forms similar in general appearance to
Eq.~(\ref{eq:elec_dist}).  Specifically, they correspond to the
\teq{\eta >1} portion of parameter space (at least for ions), so that
dissipation in the shock layer leads to a heating plus non-thermal
acceleration that ensues only at super-thermal energies in the
Maxwell-Boltzmann tail.  Evidence for such a structured distribution
can be found both in heliospheric measurements of accelerated ions near
shocks (e.g. Earth's Bow Shock: see Ellison, M\"obius \& Paschmann
1990; interplanetary shocks: see Baring et al.  1997), and also a
variety of acceleration simulations and theoretical analyses (e.g. see
Scholer, Trattner \& Kucharek 1992; Giacalone et al.  1993, Baring,
Ellison \& Jones 1993; Ellison, Baring \& Jones 1996; Kang \& Jones
1997).  The situation for relativistic shocks is less certain, due in
no small part to the absence of {\it in situ} spacecraft determinations
of particle spectra; simulation expectations for such environments are
discussed below.  In this paper, values of \teq{\eta\sim 3} appear,
being in general only weakly constrained by the fits and largely
motivated by the above physical and theoretical considerations.

While numerical evaluations of Eq.~(\ref{eq:synch_emiss}) given the
electron distribution in Eq.~(\ref{eq:elec_dist}) were performed to
obtain approximations to burst spectra, standard asymptotic limits are
readily attainable in analytic form.  The non-thermal contribution to
the synchrotron spectrum (from the truncated power-law), dominant
whenever \teq{\epsilon\gg \eta^{2+\delta}}, assumes the following 
power-law forms (derived, for example, from Eq.~[\ref{eq:synchfinal}]):
\begin{equation}
   \ndotsyn (\erg )\Bigl\vert_{\rm NT} \propto \sin\theta
   \cases{ \Bigl(\dover{\erg}{\sin\theta}\Bigr)^{-2/3}\;
           \vphantom{\biggl(},&
              $\quad \erg\ll \gammat^2 B_{\perp}/B_{\rm cr}$,\cr
           \Bigl(\dover{\erg}{\sin\theta}\Bigr)^{-(\delta +1)/2}\;
           \vphantom{\biggl(} ,&
              $\quad \erg\gg \gammat^2 B_{\perp}/B_{\rm cr}$.\cr}
 \label{eq:ntsynch_limits}
\end{equation}
Here, \teq{B_{\perp}=B\sin\theta} is the component of the field
perpendicular to the instantaneous electron motion.  The thermal
component, sampled when \teq{\epsilon\ll \eta^{2+\delta}}, necessarily
declines exponentially at photon energies well above the thermal peak,
but possesses the same power-law dependence as in the non-thermal case
for low photon energies:
\begin{equation}
   \ndotsyn (\erg )\Bigl\vert_{\rm TH} \propto \erg^{-2/3},
   \quad \erg\ll \gammat^2 B_{\perp}/B_{\rm cr}\;\; .
 \label{eq:thsynch_limits}
\end{equation}
This spectral similarity at low energies follows from the narrow nature
of the distributions being convolved with the synchrotron emissivity
functions in Eq.~(\ref{eq:synchrate}).  Accordingly, it is not possible
to distinguish between emission from thermal or power-law distributions
if spectrum measurements exist only at energies below the peak; some
other diagnostic is required.  The exponential decline above the
thermal peak possesses a somewhat complex two-domain structure; the
reader is referred to Pavlov \& Golenetskii (1986), Brainerd \&
Petrosian (1987) and Baring (1988a) for the presentation of appropriate
asymptotic forms in cases \teq{\gammat\gg 1}.

\begin{figure*}[t]
\twofigureoutpdf{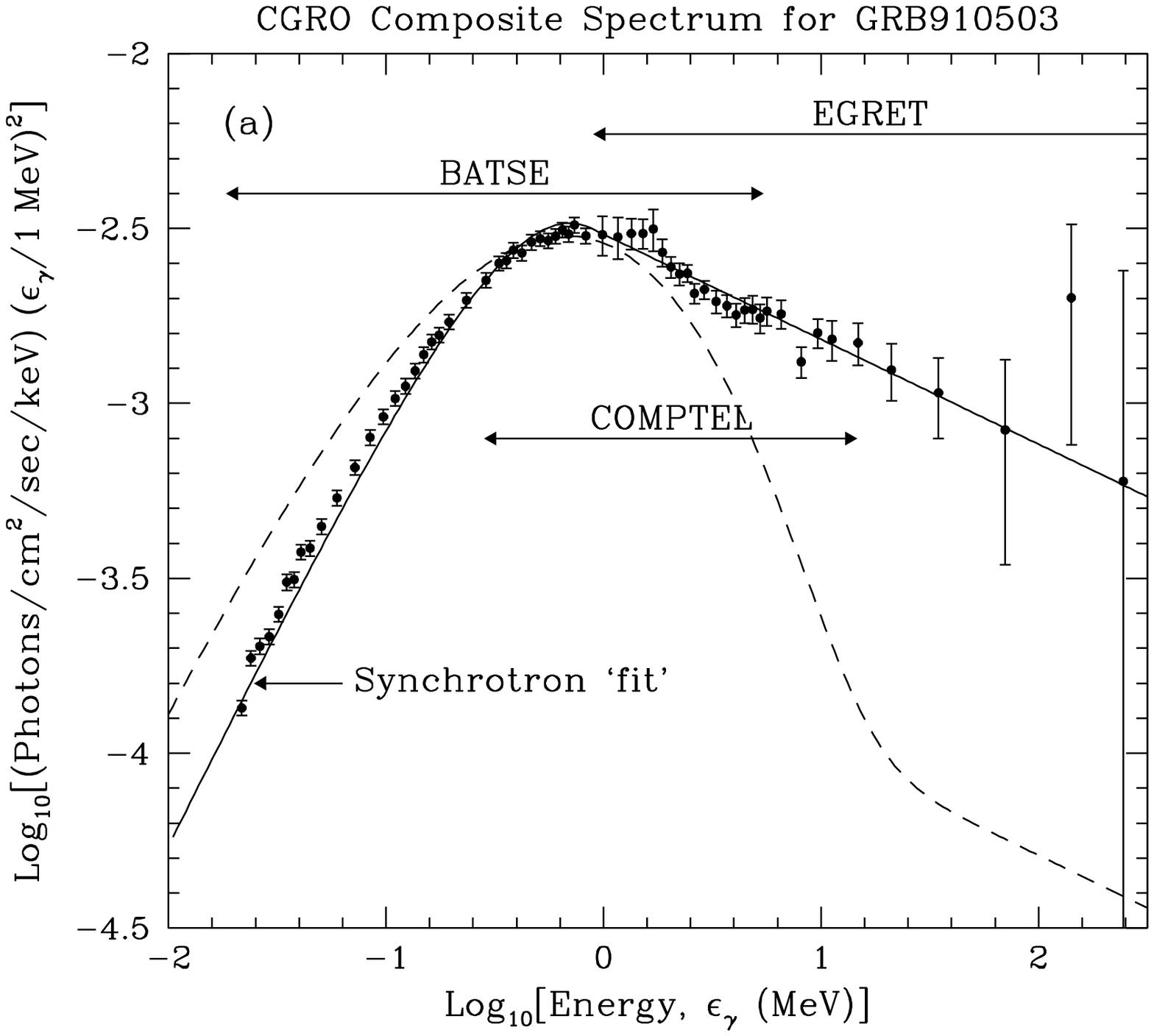}{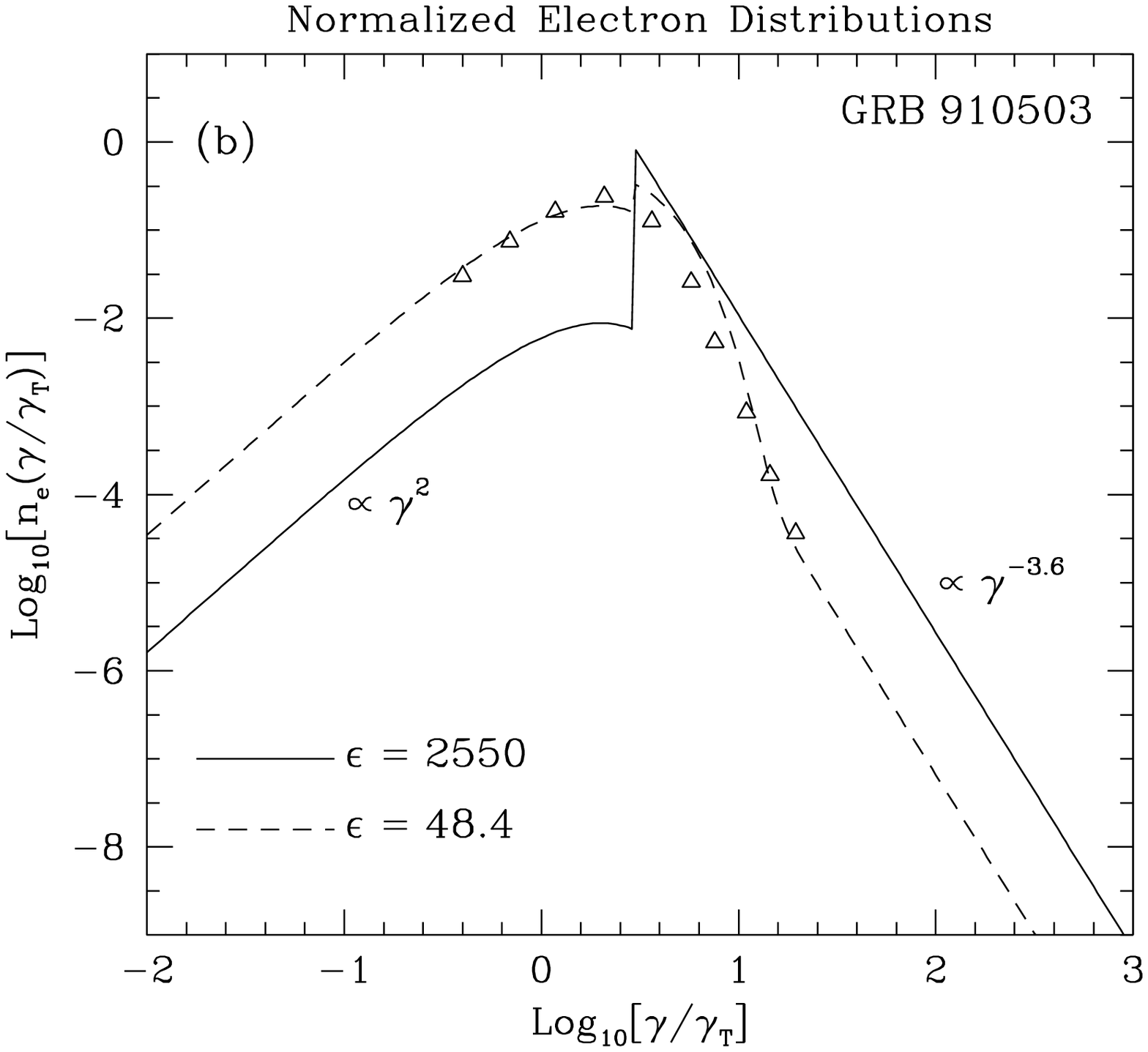}{
Photon spectra (a) from the synchrotron model and the corresponding
normalized electron distributions (b) for the bright burst GRB 910503
observed by the Compton Gamma-Ray Observatory.  The data compilation
presented in (a), adapted from Baring (1995), is a pre-publication
release of that in Schaefer et al. (1998).  The operating energy bands
of the three detecting instruments, BATSE, COMPTEL and EGRET, are as
indicated.   In panel (b), the distributions are normalized to unit
area using a scaling of the Lorentz factor in terms of \teq{\gammat}
(see text).  The solid curves correspond to the particle
distribution/resulting synchrotron spectrum pair that approximates the
observed continuum very well [labelled synchrotron 'fit' in panel (a)];
model parameters included \teq{\eta=3.0} and \teq{\epsilon\approx
2550}.  The dashed case (for \teq{\eta=3.0} and \teq{\epsilon=48.4})
illustrates a thermally-dominated situation that fails to account
for the observations.  In both cases, the non-thermal distribution
index was \teq{\delta=3.6}, the electron pitch angle was assumed to
be \teq{\theta=\pi/2}, and \teq{\Gamma\gammat^2 B_{\perp}/B_{\rm
cr}=0.90} arose from adjustments along the energy axis.  The open
triangles depict the approximate distribution obtained from the full
PIC simulation of a plane-parallel relativistic electron-positron
plasma shock  (Mace and Jones 1994; see text).
 \label{fig:synch_spec_dist} }      
\end{figure*}

It must be noted that throughout this paper, photon spectra are
generated in the shock rest frame (SRF) and then boosted in energy to
the observer's frame (OF) by a fixed Doppler factor, independent of
photon angle in the SRF.  This approximation, essentially adopted in
many models of GRB spectra, is expedient but does not accommodate the
fact that photon distributions in the SRF are inherently anisotropic
due both to the instrinsic nature of the synchrotron and inverse
Compton processes, and also to the fact that relativistic shocks
naturally generate anisotropic ions and electrons.  Notwithstanding,
this boost approximation does not dramatically distort the
angle-integrated OF spectral shape for typical SRF angular
distributions and suffices to describe the continuum for the purposes
of the discussions here.

\subsection{Inferred Electron Distributions}
 \label{sec:part_synch}

Before presentation of results, it is appropriate to address the
modeling procedure used in this paper. Approximate spectra, as opposed
to data fits, were derived visually using both log-log and semi-log
representations of spectra to increase confidence in the results.  Least
squares trials were performed in some cases, but with no addition of
insight or perceived accuracy.  This choice of a visual approach was one
of expediency, and did not compromise the scientific conclusions
presented here; the essential points are unambiguously made without the
need of statistical analysis.  This latitude is afforded by the limited
energy range of the data in the BATSE catalog: slight adjustments in
parameters used to fit the extant data for a given burst would almost
certainly have been dominated by those needed to accommodate a broader
spectral range.  For the specific cases of GRB 910503 and GRB 910601,
where EGRET data are incorporated in the study, the uncertainty in the
\teq{>100}MeV data points is a major contributor to the determination of
model parameters.  In bursts where no EGRET or COMPTEL data were used,
the ``fit'' was actually to the Band (1993) model spectral fit presented
in the Preece et al. (2000) spectroscopy catalog, not to the actual
data.  In such cases, the high energy Band model index was adopted, and
other distribution parameters adjusted.  In this sense, there was in
implicit second order incorporation of the statistics embodied in the
Band model fit.  It must be emphasized that there is no uniquely
preferred method for statistically fitting public data since it is
already a convolution of model and instrumental response. The correct
approach (as in Lloyd \& Petrosian 2000) would be to fold spectral
models here through the BATSE or EGRET detector response function, an
enterprise that rapidly becomes more ambitious when different
instrumental responses must be blended.  Such complexity seems
premature for the explorations of this paper, and is deferred until the
more refined data of Swift and GLAST becomes available.

Results for the spectral approximation of the GRB 910503 data are
presented in Figure~\ref{fig:synch_spec_dist}, together with the
multi-instrument, time-integrated CGRO data as provided by Schaefer
(private communication) prior to the public release in Schaefer et al.
(1998).  We remark that the spectral deconvolutions published in 
Schaefer et al. (1998) were made independently for each data set from 
the different CGRO instruments, applying an artificial normalization to
reconcile offsets in the different data amplitudes.  Such a
procedure is less accurate than a joint spectral deconvolution of
the entire spectral data set (see for example the analysis of GRB
990123 data in Briggs et al. 1999). Notwithstanding, the major
conclusions of this paper are robust and are not affected by such
subtleties of data analysis.

Figure~\ref{fig:synch_spec_dist} displays both synchrotron emission 
spectra for the case of pitch angles \teq{\theta =\pi/2} (the so-called 
{\it quasi-isotropic} case), and the inferred electron distribution that
generates the continuum.  For the solid curve approximate case, the overall
normalization is obviously a free parameter and was offset slightly
from the data for visual clarity.  The electron distributions were
normalized to unit area in \teq{\gamma/\gammat} space.  The solid curve
depicts the case that would be required to explain the photon
spectrum.  The non-thermal particle distribution clearly dominates the
thermal component in this ``fit.''  Its index differs from the value of
3.8 that Tavani (1996a) obtained in his analysis of this burst, a
discrepancy that is actually insignificant given the error bars in the
EGRET data.  Adjustment of the \teq{\eta} parameter to improve the
appearance of the model approximation around and above the spectral
peak led to the assumed value of \teq{\eta=3.0}, accurate to around
{15\%}.  Notwithstanding, there appears to be some spectral
structure around the peak that cannot be explained by the simple
particle distribution in Eq.~(\ref{eq:elec_dist}).  The positioning
of the peak in energy produced the fit value
\teq{\Gamma\gammat^2 B_{\perp}/B_{\rm cr}=0.90}, where \teq{\Gamma}
is the Lorentz factor associated with the bulk motion of the GRB
expansion.  Here, \teq{B} is interpreted as the magnetic field in the
burst expansion frame, i.e., not in the ISM or shock rest frames.

The parameter \teq{\epsilon} describing the relative normalization of
the non-thermal and thermal components was found to be
\teq{\epsilon\approx 2550} (\teq{\pm}{10\%}), a strikingly large value
that is not addressed explicitly in detail by Tavani (1996a).  A
discussion of the implications of an inferred particle distribution with
such a large \teq{\epsilon} follows just below.  A similar property is
obtained for synchrotron model fits for BATSE bursts GRB 940619 (trigger
number 3035\_6) and GRB 940817 (trigger number 3128\_5) from the Preece
et al. (2000) catalog, and also for GRB 910601 and GRB 910814 whose
composite spectra are presented in Schaefer et al. (1998).   The spectra
for bursts GRB 930131 and GRB 940217 that possess EGRET detections also
suggest a similar result, though the lack of a composite broad-band
multi-instrument spectrum for each in the literature precluded a
detailed analysis.  The parameter regime of \teq{\epsilon\gg 1} is
expected to be pervasive for BATSE bursts, due to the smoothness of
their circumpeak continua.   For comparative purposes, another
distribution and its resultant synchrotron is illustrated in
Figure~\ref{fig:synch_spec_dist} as dashed curves.  The distribution was
deliberately chosen to possess a somewhat dominant thermal component,
motivated by expectations from shock acceleration theory, and clearly
yields a spectrum that grossly deviates from the observations both well
above and well below the peak.  While these deductions are obtained
using a time-integrated spectrum, such a conclusion must apply either
for most of the burst duration, or at sub-intervals when the burst was
at its brightest.

For the synchrotron mechanism to be a viable explanation of
GRB spectra, the underlying particle distribution must be physically
realistic.   Here, expectations of particle distributions from shock
acceleration theory are the predominant focus, partly because this was
the preferred acceleration mechanism in Tavani (1996a,b), and largely
because this is the underlying energization assumed in most burst
models.  In a nutshell, both theory and observation pertaining to
non-relativistic shocks strongly indicate \teq{\eta > 1} and
\teq{\epsilon\gg 100} cases.  Note that \teq{\epsilon/\eta^4} is
essentially the parameter that demarcates the relative normalization of
the non-thermal and thermal components in Equation~(\ref{eq:elec_dist}).

Applications of shock acceleration in the heliosphere and at supernova
remnant shells more or less mandate these parameter regimes, since the
seed pool of particles that can be subjected to acceleration are
thermal and moreover plentiful.  In heliospheric environs, the source
is the cold solar wind, and there is little evidence of significant
pre-heating that is not connected to the shock system under study.  The
{\it in situ} spacecraft measurements exhibit strong heating at shocks
and the generation of dominant thermal ions in the downstream regions,
corresponding to \teq{\epsilon/\eta^4 <1} cases.   Data taken at the
high sonic Mach number Earth's bow shock (e.g. see Ellison, M\"obius \&
Paschmann 1990; Scholer, Trattner \& Kucharek 1992) and low sonic and
Afv\'enic Mach number interplanetary shocks (e.g. see Gosling et al.
1981; Baring et al. 1997) clearly support this assertion.   In
supernova remnants, direct particle measurements are, or course,
impossible, and inferences can only be made from radiative signatures.
While electron temperatures and densities can be estimated from X-ray
signals and molecular emission in the IR/optical bands, the data cannot
yet clearly discriminate \teq{\epsilon/\eta^4 \ll 1} and
\teq{\epsilon/\eta^4 \gg 1} situations.

From a theoretical perspective, diffusive shock acceleration naturally
draws particles from the thermal pool via transport in shock-associated
turbulence.  Whether this be first or second order in momentum
transport, thermal and non-thermal particles are subjected to similar
diffusion, at least as far as ions are concerned, so that
\teq{\epsilon/\eta^4 >1} circumstances would require virtually complete
suppression of thermalization and a 100\% efficiency of acceleration out
of the thermal pool.  Such a scenario would require unusually anomalous
spatial diffusion coefficients \teq{\kappa (p)} (i.e. extremely small)
at thermal momenta, or perhaps strange and non-dissipative
electrodynamic properties in the shock layer.  Note that \teq{\kappa
(p)} is more likely to increase at low energies due to the possibility
of wave damping by thermal populations (e.g. see Forman, Jokipii \&
Owens 1974, for the relationship between \teq{\kappa (p)} and wave power
spectra within the confines of quasi-linear theory).  Without the
assumption of anomalous forms for \teq{\kappa (p)}, Monte Carlo
transport simulations (e.g. see Baring, Ellison \& Jones 1993; Ellison,
Baring \& Jones 1996) clearly indicate a downstream thermal population
that dominates the extended non-thermal distribution in number,
corresponding to \teq{\epsilon/\eta^4 < 1}.  The ``injection'' of ions
from the thermal pool is extremely efficient, but not incredibly so. 
Hybrid plasma simulations (e.g. see Trattner \& Scholer 1991, 1993;
Scholer, Trattner \& Kucharek 1992; Giacalone et al. 1992, 1993; Liewer,
Goldstein \& Omidi 1993) of wave generation and ion transport near
non-relativistic shocks, which treat the electron component as a
background fluid and therefore address only some of the electrodynamic
aspects of the plasma in the shock layer, produce similar
\teq{\epsilon/\eta^4 <1} distributions.

These indications are strongly suggestive, but not decisive for the
gamma-ray burst problem, since they largely don't focus on electron
properties and equally importantly do not explore acceleration at
relativistic shocks.  In situ measurements of non-thermal electrons at
shocks are limited; examples include recent Geotail observations of
interplanetary shocks (Shimada, et al. 1999; Terasawa et al., 2001).
Accordingly, discussion is dependent on simulational information, as is
the case for relativistic shocks, and the literature is sparse on such
a subject.  Monte Carlo simulations of ion acceleration at relativistic
shocks (e.g. Ellison, Jones \& Reynolds 1990; Baring 1999;
Ellison \& Double 2002) generate the same injection properties as do
non-relativistic shocks, i.e.  \teq{\epsilon/\eta^4 < 1}.

Plasma simulations of relativistic shocks are similarly sparsely
studied.  Gallant et al. (1992) and Hoshino et al.  (1992) presented
results from one-dimensional (1D) particle-in-cell (PIC) simulations of
electron-positron and electron-positron-proton plasmas in
ultrarelativistic flows with perpendicular shocks, where the mean field
was orthogonal to the shock normal.  Such simulations treat the full
electrodynamics of such plasmas, but are severely limited by CPU
constraints.  Accordingly, they tend to use unrealistically small
proton to electron mass ratios, and model relatively small spatial
scales.  These constraints tend to mask the production of some wave
modes and limit consideration to particles with energies at most a few
to ten times thermal.  The context for these investigations was pulsar
wind flows, and they revealed no acceleration in pure \teq{e^{\pm}}
plasmas.  With the addition of protons in significant abundances,
Hoshino et al. (1992) demonstrated that non-thermal positrons (and not
electrons) could be created, as the ions generated
elliptically-polarized magnetosonic modes that resonantly interacted
with the positrons.  

Mace and Jones (1994, unpublished; F. Jones, private communication)
explored a 1D PIC simulation similar to Gallant et al. (1992) for an
ultrarelativistic \teq{e^{\pm}} plasma shock of arbitrary field
obliquity, and found that electron/positron acceleration arose in
quasi-parallel scenarios, but disappeared when the field obliquity to
the shock normal increased above around \teq{\Theta_{\rm BN}\sim 30}
degrees, concurring with zero acceleration in the earlier results for
perpendicular shocks.  A scaled version of their parallel shock
\teq{e^{\pm}} distribution (binned in energy) is depicted in
Fig.~\ref{fig:synch_spec_dist}, and is fairly proximate to the
\teq{\eta=3}, \teq{\epsilon =48.4} case.  A notable feature of the
simulation data is the very limited range of energies accessible to the
PIC technique, imposed by the severe CPU limitations for such complex
codes.  Some interesting recent PIC simulations results have been
presented by Shimada \& Hoshino (2000) and Hoshino \& Shimada (2002) for
electron-proton plasmas containing mildly-relativistic shocks, where
suprathermal power-law electrons are generated and the distribution
closely resembles the PIC simulation data depicted in
Fig.~\ref{fig:synch_spec_dist}.  Yet, as with the Mace and Jones
results, the complete distribution reveals that the power-law emerges
only well into the Maxwellian tail, corresponding to the
\teq{\epsilon/\eta^{2+\delta} \lesssim 0.1} situation elicited in the
Monte Carlo simulations.

It is noted in passing that the one-dimensional nature of these
simulations potentially imposes an important absence of critical physics
to the acceleration problem: Jokipii, K\'ota \& Giacalone (1993) and
Jones, Jokipii and Baring (1998) demonstrate that particle diffusion
across field lines is eliminated when simulations restrict the spatial
dimensions to less than three.  The implication of this result for the
\teq{e^{\pm}} shocks is a cessation of acceleration when thermal
particles cannot be transported across fields to return to the shock
from the downstream side; Baring, Ellison \& Jones (1993) demonstrated
that for high Mach number, non-relativistic ion shocks, a quenching of
acceleration through this effect would arise at around \teq{\Theta_{\rm
BN}\sim 30^\circ}, a result that matched the behavior seen in the
relativistic PIC simulations of Mace and Jones.  Moreover this result
is consistent with the absence of acceleration in non-relativistic 1D
hybrid plasma simulations applied to the quasi-perpendicular
heliospheric termination shock (e.g. Liewer, Goldstein \& Omidi 1993;
Kucharek \& Scholer 1995), a site that is believed to generate the
non-thermal anomalous cosmic rays.  The essential loss of physics in
simulations of restricted dimensionality is clearly crucial to the correct
modeling and interpretation of particle acceleration in gamma-ray bursts.

The principal conclusion of this discussion is that there is little
observational or simulational evidence for quasi-isotropic electron
distributions in the environs of shocks that would render a synchrotron
spectrum commensurate with the GRB910503 data in particular and typical
burst spectra in general.  

\subsection{The Low Energy Spectral Index Issue}
 \label{sec:alphalow_synch}

An issue that has already been substantially discussed in the literature
is whether the synchrotron model is compatible with spectral slopes
below the burst peak.  For GRB 910503, Fig.~\ref{fig:synch_spec_dist}
clearly indicates that there is no conflict between the data and the
model.  The fixed value of \teq{\alpha =-2/3} for the low energy index
in Eqs.~\ref{eq:ntsynch_limits} and~\ref{eq:thsynch_limits} suggests
that this is not the situation for all bursts, a fact that was
demonstrated by Preece et al. (1998; see also Preece et al. 2000; 2002).
Synchrotron theory distinctly predicts this canonical \teq{-2/3} index
at low energies for any isotropic electron distribution that has a
characteristic Lorentz factor \teq{\gamma_c} as its lower scale. 
Correlations between Lorentz factors and pitch angles can alter this
property, as discussed in Section~\ref{sec:pitch_synch} below.  Any
index greater than this value cannot be accommodated by quasi-isotropic
distributions; indices less than this are easily described by a
steepening particle distribution above \teq{\gamma_c}. Preece et al.'s
(2000) display of the distribution of low energy indices clearly
indicated a conflict, as illustrated in Fig.~\ref{fig:BATSE_low_index}. 
There the low energy indices for various empirical fitting models used
in the BATSE spectroscopy catalog are binned in a histogram, with the
permitted range of \teq{\alpha <-2/3} clearly marked.  The most common
fitting model used in the catalog was the popular Band (1993) GRB model
for the observed flux \teq{f_{\gamma}(\erg )} in photons/cm$^2$/sec:
\begin{equation}
   f_{\gamma}(\erg ) \; =\; f_0
   \cases{ \biggl( \dover{\erg}{\erg_0}\biggr)^{\alpha}\,
           \exp\biggl( -\dover{\erg}{\erg_0}\biggr) ,&
              $\quad \erg\leq (\alpha -\beta )\erg_0$,\cr
           \biggl( \dover{\erg}{\erg_0}\biggr)^{\beta}\,
           \exp ( \beta -\alpha ) ,&
              $\quad \erg > (\alpha -\beta )\erg_0$.\cr}
 \label{eq:Bandmodel}
\end{equation}
Most bursts possess values \teq{\alpha <0} and \teq{\beta <0}.  In
folding the Band model, broken power-law models and others through the
BATSE detector response, Preece et al. (2000) found that the Band model
performed best in roughly 2/3 of the burst sample.  Note that the index
\teq{\alpha} recorded in the catalog is the asymptotic low energy index
(Preece et al. 2000), and does not exactly match the index local to the
lowest energy \teq{\sim 25}keV in the BATSE window.  The inferred values
of \teq{\alpha} are subject to instrumental selection effects,
particularly when the peak energy is low.  Hence, there is the
possibility that the \teq{\alpha} distribution might move to slightly
higher \teq{\alpha} when the detection band moves to lower energy, the
situation for Swift.  Note that such an effect was observed by Preece et
al. (1998) when applying a correction to accommodate the energy of the
BATSE window.  A depiction of the \teq{\alpha} histogram for
time-integrated spectra in the BATSE spectroscopy catalog is given in
Lloyd \& Petrosian (2000); it exhibits a similar distribution, as
expected.

\figureoutpdf{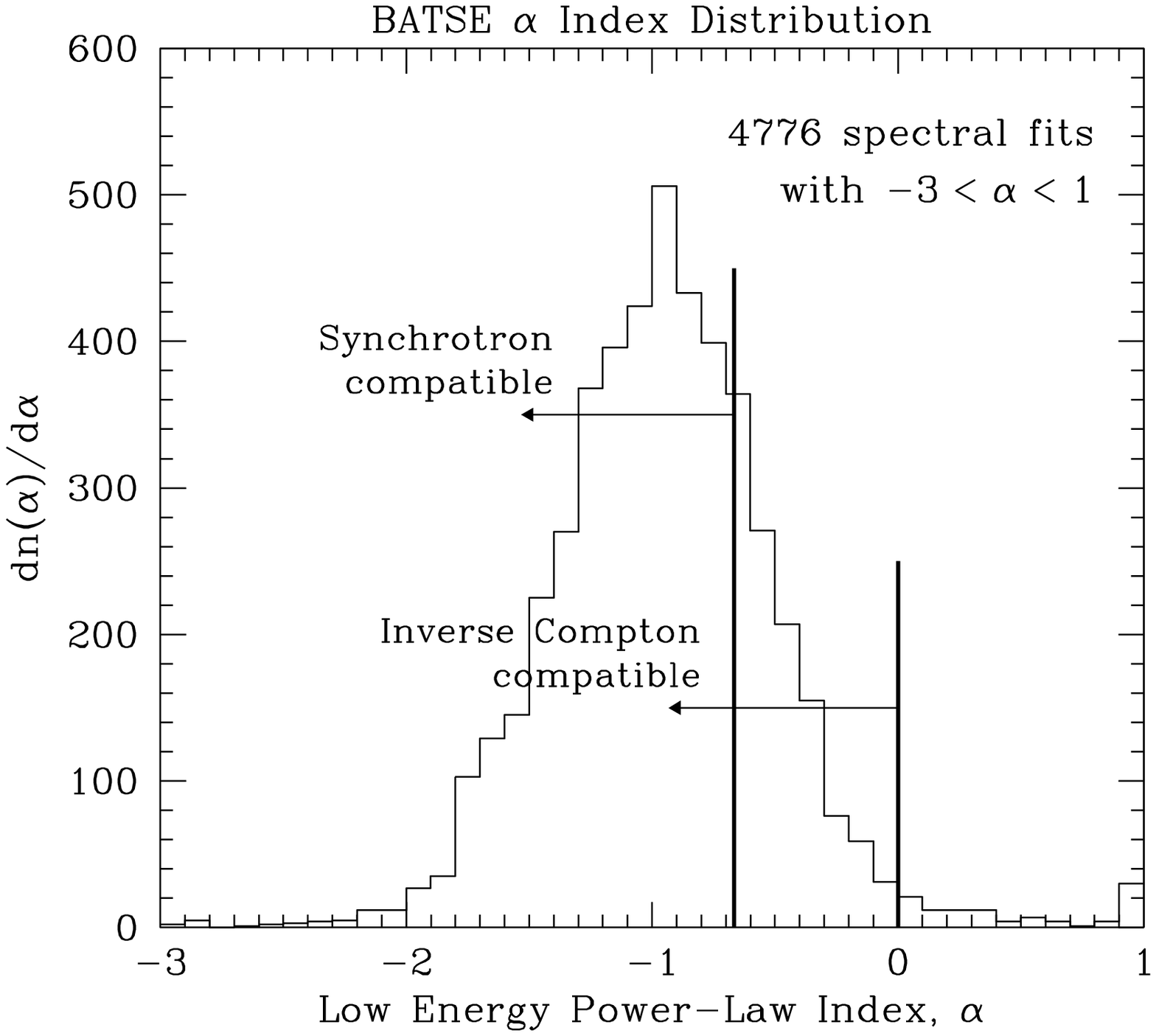}{3.4}{0.0}{-0.2}{
The low energy power-law index \teq{\alpha} distribution presented in
Figure~7 of the BATSE spectroscopy catalog of Preece et al. (2000).
The index is obtained by Band model or broken power-law model fits to
burst spectra (see text).  The public database consists of some 5500
spectra from sub-intervals in 156 bursts, for which a subset of 4776
spectra possess well-determined \teq{\alpha}.  Domains where a
quasi-isotropic synchrotron (\teq{\alpha <-2/3}; 75.9\% of spectra) or
inverse Compton (\teq{\alpha <0}; 97.8\% of spectra) models can account
for the observed low-energy index are as indicated.
 \label{fig:BATSE_low_index}} 

The quasi-isotropic synchrotron model clearly cannot account for nearly
25\% of sample burst spectra.  This is true even for pathological
choices of electron energy distributions: from the form in
Eq.~(\ref{eq:synch_emiss}), it is easily shown that the spectral index
\teq{\alpha = -\partial\log_{10} \ndotsyn (\erg )/d\log_{10}\erg} is a
monotonically increasing function of energy \teq{\erg} for any electron
distribution, due to the broad and smooth nature of the synchrotron
emissivities in Eq.~(\ref{eq:synchrate}).  While this is disturbing, it
may be more of an indication of problems with assumptions underpinning
the \teq{\alpha <-2/3} bound than a suggestion that synchrotron
radiation is not responsible for GRB prompt emission. For instance,
relaxing an adherence to isotropy or \teq{n_{\theta}\propto \delta
(\theta -\theta_c)} distributions can potentially alter the spectral
slope.  It is demonstrated in Section~\ref{sec:anis_synch} below that
for an \teq{n_{\theta}} independent of \teq{\gamma}, the spectrum below
the peak cannot be flattened beyond \teq{\alpha =-2/3} cases, and in
fact steepens. However, introducing a \teq{\gamma}-dependent angular
distribution can suitably ease the \teq{\alpha} bound; this is
exemplified in the focus on pitch angle synchrotron emission in
Section~\ref{sec:pitch_synch}.

An alternative recourse is synchrotron self-absorption, i.e. relaxation
of an optically thin assumption.  Given the theoretical expectation of
extremely flat spectra below the \teq{\tau_{\rm syn}=1} break
(\teq{\erg\ndotsyn \propto\erg^{5/2}} for power-law electrons and a
\teq{\erg\ndotsyn \propto\erg^2} Raleigh-Jeans tail for thermal
\teq{e^-}: see, e.g., Rybicki \& Lightman 1979), different groups have
invoked such a mechanism to account for \teq{\alpha > -2/3} cases in the
BATSE catalog (e.g. Crider \& Liang 1999; Granot, Piran \& Sari 2000). 
Pushing the absorption break to gamma-ray energies requires extreme
physical conditions (\teq{n_e\sim 10^8}cm$^{-3}$, \teq{B\sim
10^8}Gauss), as noted by Lloyd \& Petrosian (2000), and can be inferred
from Eq.~(\ref{eq:selfabs_energy}).  These parameter regimes would be
more extreme in the absence of the strong Doppler boosting associated
with GRB outflows.  Equipartition is usually invoked to justify such
high fields, however a widely-accepted and well-understood mechanism for
the generation of such fields remains elusive.  A convenient resolution
of this issue is afforded by the proposition of Panaitescu \& Meszaros
(2000), namely that the prompt emission is an inverse Compton image of a
self-absorbed synchrotron spectrum.  The concomitant moving of the
absorption break energy \teq{\erg_t} to below the optical band lowers
the required values of \teq{B} and \teq{n_e} to less extraordinary
ranges.  The X-ray spectral indices in the scenario of Panaitescu \& 
Meszaros (2000; see also Liang, Boettcher \& Kocevski 2003) can 
comfortably accommodate the \teq{\alpha =0} domain, as discussed in
Section~\ref{sec:alphalow_comp}.

An additional concern is that the turnover energy \teq{\erg_t} (in the
GRB expansion frame) satisfying \teq{\tau_{\rm syn}(\erg_t)=1} is a
function of a number of environmental parameters.  Equation~(6.53) of
Rybicki \& Lightman (1979) for the optical depth \teq{\tau_{\rm syn}}
for self-absorption can be manipulated to arrive at the equation for
\teq{\erg_t}.  This result can be found in many papers; here Eq.~(7) of
Baring (1988b) is adapted to accommodate the specific form for the
non-thermal portion of the distribution in Eq.~(\ref{eq:elec_dist}):
\begin{equation}
   \erg_t^{(\delta+4)/2}\; =\; \phi(\delta )\; \dover{\taut}{\fsc}\;
   \biggl( \dover{B_{\perp}}{B_{\rm cr}} \biggr)^{(\delta+2)/2}\; 
   (\eta \gammat )^{\delta-1}\;\; ,
 \label{eq:selfabs_energy}
\end{equation}
for
\begin{equation}
   \phi(\delta )\; =\; \dover{3^{(\delta+3)/2}}{16}\, (\delta -1)\;
   \Gamma \biggl( \dover{\delta}{4}+\dover{1}{6} \biggr)\, 
   \Gamma \biggl( \dover{\delta}{4}+\dover{11}{6} \biggr)\;\; . 
 \label{eq:phidef}
\end{equation}
Here, \teq{\taut} is the Thomson optical depth of the emission region,
and \teq{B_{\perp}=B\sin\theta}, as before.  Note that \teq{\Gamma (x)}
represents the Gamma function in Eq.~(\ref{eq:phidef}).  The density and
size of the absorbing medium are incorporated in \teq{\taut}.  These
parameters possess implicit dependence on the bulk Lorentz factor of the
flow, with different evolutionary dependences for adiabatic and
radiative expansion cases (e.g., see Dermer, Chiang \& B\"ottcher 1999).
  If Eq.~(\ref{eq:selfabs_energy}) is interpreted in the frame of the
burst outflow, then \teq{\erg_t} is in the EUV/soft X-ray band for
typical burst Lorentz factors.  Given the number of parameters (five in
all) contributing to the computation of the self-absorption optical
depth, it is difficult to comprehend how \teq{\erg_t} can be fine-tuned
to the BATSE band in the observer's frame; such a conspiracy is an issue
also for optically thin synchrotron emission, whose peak depends on
three parameters, \teq{\Gamma}, \teq{\gammat} and \teq{B_{\perp}}.

A strong test of the self-absorption proposition could be offered by
Swift, which should determine whether or not BATSE was fully capable of
determining the ``asymptotic'' values of \teq{\alpha}, especially for
those bursts with peaks around 100 keV or lower.  The Gamma-Ray Burst
Monitor (GBM) on GLAST will provide an additional broad-band probe.
Should the BATSE \teq{\alpha} distribution be mirrored by the Swift and
GBM ones, synchrotron self-absorption in the gamma-ray band could then
only be operating in a small minority of bursts, rendering it less
germane to the spectral shape discussion, except perhaps for synchrotron
self-Compton models.

\subsection{Influences of Anisotropy}
 \label{sec:anis_synch}

The foregoing discussion has focused on the case of synchrotron
emission from quasi-isotropic or large pitch angle electron
distributions.  Distributing the angles of electrons in a non-uniform
manner can alter the spectral index of the emitted radiation.  A prime
example of this was the flattening incurred in distributing magnetic
pair creation turnovers in neutron star environments, probed by Baring
(1990).  Here, obviously a distribution skewed towards small pitch
angles is a potential candidate for modifying \teq{\alpha}.  Such a
distribution is not at all contrived:  synchrotron cooling can produce
it quite naturally (e.g. see Brainerd \& Lamb 1987).  However, as
alluded to above, it is now demonstrated that such distributions, if
uncorrelated with Lorentz factor, yield only spectral steepening, and
so cannot resolve the low energy index problem.

The spectral behavior can adequately be demonstrated using the
non-thermal portion of Eq.~(\ref{eq:elec_dist}), which for the purposes
of this discussion will be written in the form  \teq{n_e(\gamma
,\;\theta ) = f_{\theta}\, n_e\, (\delta -1)\, \gammin^{(\delta -1)}\,
\gamma^{-\delta}} for \teq{\gamma\geq\gammin} and is zero otherwise.
Here \teq{\gammin =\eta\gammat}.  If \teq{\eb = B/B_{\rm cr}} is
defined as the dimensionless cyclotron energy, then the differential
rate of production of synchrotron photons in Eq.~(\ref{eq:synch_emiss})
can be quickly manipulated by reversing the integrations over
\teq{\gamma} and \teq{y}.  The result is
\begin{equation}
  \ndotsyn (\erg )\; =\; \Ndotsyn \,\dover{\delta - 1}{\delta + 1}
    \int_0^{\infty} dy \, K_{5/3}(y)\,  \xi^{-(\delta +1)/2},
 \label{eq:synchfinal}
\end{equation}
where
\begin{equation}
   \Ndotsyn\; =\; \dover{\fsc}{\pi\sqrt{3}}\,
   \dover{c}{\lambar}\,\dover{n_e}{\gammin^2}\,  f_{\theta}\sin\theta
 \label{eq:Ndotsyn_def}
\end{equation}
is the synchrotron emissivity scale, and
\begin{equation}
  \xi\; =\; \max \biggl\{ 1,\;
   \dover{2\erg }{3y \eb\gammin^2\sin\theta} \,\biggr\}\quad .
 \label{eq:chi_def}
\end{equation}
This form for the synchrotron rate will be used later in the
discussions on synchrotron self-Compton emission.  By adding
and subtracting suitable forms like Eq.~(\ref{eq:synchfinal}) with
different choices of \teq{\gammin}, a general expression for the
synchrotron emission from a piece-wise continuous power-law
distribution of electrons can be obtained.  Defining the photon
spectral index to be \teq{\alpha = -\partial\log_{10} \ndotsyn (\erg
)/d\log_{10}\erg}, it is easily shown that \teq{\alpha} is a
monotonically increasing function of energy \teq{\erg} for any electron
distribution with a low \teq{\gamma} cutoff, using piecewise continuous
power-laws to successively refine approximations to a smooth
distribution.

Eq.~(\ref{eq:synchfinal}) possesses two readily identifiable asymptotic
limits, namely when \teq{\erg \ll \eb\gammin^2\sin\theta} and \teq{\erg
\gg \eb\gammin^2\sin\theta}.  These lead to the forms given in
Eq.~(\ref{eq:ntsynch_limits}).  Now consider an angular distribution
defined over the range \teq{\thetamin\leq\theta\leq \thetamax} for
\teq{\thetamin\ll 1}.   The superposition of spectral forms in
Eq.~(\ref{eq:ntsynch_limits}) then immediately indicates that such an
angular distribution must generate a spectrum with asymptotic forms
\teq{\erg^{-2/3}} at \teq{\erg \ll \eb\gammin^2\sin\thetamin}, and
\teq{\erg^{-(\delta -1)/2}} when \teq{\erg \gg
\eb\gammin^2\sin\thetamax}.  The spectrum in between depends on how
rapidly \teq{f_{\theta}} varies with angle, and can be either of slope
\teq{2/3} or steeper, but never flatter than \teq{\erg^{-2/3}} and
never steeper than \teq{\erg^{-(\delta -1)/2}}.  This follows from the
convolution over angles of spectra that are monotonically declining in
energy necessarily being also monotonically declining in \teq{\erg}.
This assertion holds even for \teq{f_{\theta}} that is strongly peaked
towards \teq{\thetamin}, and monotonically declining with
\teq{\theta}.  Hence, angular distributions with \teq{\theta}
uncorrelated with \teq{\gamma} cannot ameliorate the low energy index
issue due to the spectral shape implied by Eq.~(\ref{eq:synchfinal}).

\subsection{Pitch Angle Synchrotron}
 \label{sec:pitch_synch}

A more viable possibility for flattening the low energy index is when
there is a correlation between the pitch angle \teq{\theta} and the
Lorentz factor \teq{\gamma} of charged particles. This provides a
distinct kinematic difference from the foregoing considerations of
synchrotron radiation from quasi-isotropic electron distributions.  The
archetypical example for consideration here is so-called {\it
pitch-angle synchrotron} radiation, a mechanism explored in detail by
Epstein (1973; for applications, see Epstein \& Petrosian 1973). This
mechanism corresponds to synchrotron emission from particles possessing
very small pitch angles \teq{\theta\ll 1/\gamma}. The
polarization-averaged emissivity derived by Epstein (1973) in this
limit, for electrons in a uniform magnetic field, can be written in the
form
%
%
\begin{equation}
   \ndotpas\; =\; \pi\sqrt{3}\; \gamma^2\theta^2\, {\dot N}_{\theta}\, 
   \Bigl( 1-2\Psi + 2 \Psi^2 \Bigr)\;\; , \;\; 
   \Psi\; =\; \dover{\erg B_{cr}}{2\gamma B}\;\; .
 \label{eq:synch_pas_rate}
\end{equation}
This form is simply obtained from the flux emissivity in Equation~(10)
of Epstein (1973), and is applicable for \teq{\erg\lesssim \erg_c
/\gamma}; at higher energies, the usual synchrotron formula in
Eq.~(\ref{eq:synch_emiss}) is operable if \teq{\gamma\theta\gtrsim 1}.
Note that this process exhibits strong circular polarization due to the
near coalignment of particles with the field, though there is
significant linear polarization in narrow bands of photon energy.
Observe that \teq{\Psi\equiv 3\gamma\erg/(4\erg_c)\lesssim 1} defines
the intrinsic energy scale of pitch angle synchrotron (PAS) emission. 
Eq.~(\ref{eq:synch_pas_rate}) is virtually independent of \teq{\erg} for
domains \teq{\Psi\ll 1}, thereby defining an extremely flat spectrum. 
This property is a direct consequence of the kinematic coupling
\teq{\theta\sim 1/\gamma}, yielding a dependence \teq{\erg\sim \gamma
B/B_{cr}} that contrasts the isotropic synchrotron dependence
\teq{\erg\sim \gamma^2\theta B/B_{cr}}. This weaker dependence on
\teq{\gamma} immediately implies an inference of stronger source fields
in order to move the characteristic PAS photon energy into the BATSE
window.

The spectrum drops dramatically above \teq{\Psi\sim 1} if
\teq{\gamma\theta\ll 1}, otherwise it merges into the \teq{\erg^{-2/3}}
synchrotron form if \teq{\gamma\theta\gtrsim 1}. These shapes imply (i)
that prompt GRB spectra that are flatter than \teq{\erg^0} below the
peak (i.e. around 98\% of bursts: see Figure~\ref{fig:BATSE_low_index})
can be easily accommodated by a pitch angle synchrotron scenario (just
like the inverse Compton scattering case considered in
Section~\ref{sec:ICformalism} below), and (ii) accordingly that a
variety of GRB spectral shapes can be accurately modeled due to the
characteristically narrow peaking of the PAS emissivity in
Eq.~(\ref{eq:synch_pas_rate}).  These were conclusions of Lloyd-Ronning
and Petrosian (2002), who explored fits to BATSE bursts for pitch angle
synchrotron models, and addressed correlations between spectral
parameters and temporal properties. Hence, pitch angle synchrotron is an
attractively versatile spectroscopic possibility for bursts. Note that
the mechanism of ``jitter'' radiation in turbulent magnetic fields
(Medvedev 2000) possesses similar spectral properties due to the
intrinsic \teq{\theta\sim 1/\gamma} kinematic coupling, thereby offering
an alternative possibility of similar promise.

Pitch angle synchrotron might arise when electrons stochastically
increase their pitch angle from virtually zero during an acceleration
process. For it to provide the major contribution, a prevalence of small
pitch angles must be strongly favored, i.e. the pitch angles
\teq{\theta\gtrsim 1/\gamma} must not be significantly populated, so
that normal synchrotron emission does not dominate.  It is not clear how
this might come about, though rapid cooling and inefficient diffusion
transverse to the field is an interesting possibility (e.g. see
Lloyd-Ronning and Petrosian 2002). In addition, small pitch angles must
be prepared in the first instance, which suggests a coherent
electrodynamic mode of acceleration as opposed to diffusive one. While
PAS may afford a means of enhancing the non-thermal radiative signal
relative to the thermal component, a detailed exploration of PAS
spectral shapes and polarization for thermal and non-thermal
distributions in the context of diffusive acceleration will be the
subject of future work.

\subsection{Cooling Issues}
 \label{sec:cooling}

While detailed exploration of cooling effects is beyond the scope of
this paper, a brief discussion is appropriate.  Cooling is known to
strongly influence spectral shape in quasi-isotropic synchrotron
scenarios (e.g. Ghisellini, Celotti \& Lazzati 2000), introducing a
second ``line of death'' with a value of \teq{\alpha = -3/2}.  This
presents problems for models invoking strong cooling in generating
spectra like those seen by CGRO. Such models are most applicable to GRB
environments where the emission regions are somewhat remote from the
acceleration ones.  This is because even though synchrotron radiation is
a remarkably efficient process, diffusive acceleration at shocks is even
more so.  Accordingly, coincident acceleration and emission zones of the
same physical scale can exhibit prolific acceleration up to some maximum
Lorentz factor \teq{\gammax}, before radiative cooling in the magnetic
field causes cessation of acceleration.  The spectral turnover that
marks this cessation is quasi-exponential in nature, and have not been
observed in burst spectra to date.  This combination
acceleration/radiation scenario is widely envisaged in the contexts of
X-ray and gamma-ray emission from supernova remnants and blazars. 
Whenever multiple spatial or temporal scales are sampled, convolution of
spectral contributions can blur the turnover into a broad distribution,
essentially the case explored by Ghisellini, Celotti \& Lazzati (2000).

The rate of acceleration scales as the inverse gyrofrequency, which has
a weaker dependence on electron Lorentz factor \teq{\gamma} than the
synchrotron cooling rate.  Specifically, the acceleration timescale (in
seconds) can be approximated by \teq{\tau_{\hbox{\sevenrm acc}}\sim 0.1
f_{\hbox{\sevenrm rel}} E_{\hbox{\sevenrm TeV}}/(\beta_{\hbox{\sevenrm
sh}}^2\, B)}, where \teq{E_{\hbox{\sevenrm TeV}}} is the energy of the
electrons in units of TeV, \teq{\beta_{\hbox{\sevenrm sh}}} is the shock
speed in units of \teq{c}, \teq{B} is the mean magnetic field strength
(measured in Gauss), and \teq{f_{\hbox{\sevenrm rel}}} is a factor of
order unity that accounts for corrections to the diffusive timescale
appropriate for relativistic shocks. The synchrotron cooling timescale
(also in seconds) is \teq{\tau_{\hbox{\sevenrm syn}}\sim 300/(B^2\,
E_{\hbox{\sevenrm TeV}})}. Various versions of these formulae are widely
available, though a specific discussion of them in the context of
blazars is in Baring (2002); they are effectively applicable in the
comoving frame of the burst outflow, as opposed to the observer's frame.
 The electron energy and the magnetic field are coupled by placing the
synchrotron \teq{\nu F_{\nu}} peak in the soft gamma-ray band, i.e.
setting \teq{\Gamma\,\gamma^2B\sim 10^{14}}Gauss. Here \teq{\Gamma\sim
10^2}--\teq{10^3} is the bulk Lorentz factor of the outflow.  These
quickly lead to the ratio estimate
\begin{equation}
   \dover{\tau_{\hbox{\sevenrm syn}}}{\tau_{\hbox{\sevenrm acc}}}\; \sim\;
   10^{16}\, \dover{\beta_{\hbox{\sevenrm sh}}^2}{f_{\hbox{\sevenrm rel}}}\;
   \dover{1}{\gamma^2B}\;\sim\; 100\,\Gamma\; 
   \dover{\beta_{\hbox{\sevenrm sh}}^2}{f_{\hbox{\sevenrm rel}}}\quad .
 \label{eq:cooling_ratio}
\end{equation}
Acceleration is clearly much more rapid than synchrotron cooling for
typical \teq{\Gamma\gtrsim 10^2}, a property that changes only when
considering electrons that emit above about 10 GeV, i.e. for
\teq{\Gamma\,\gamma^2B\gtrsim 10^{18}}Gauss.  In the special case of
\teq{\Gamma\sim 1}, the result in Eq.~(\ref{eq:cooling_ratio})
corresponds to the well-known synchrotron cooling cutoff property of the
Crab pulsar wind nebula (e.g. see de Jager et al., 1996), namely that
the spectral turnover occurs at around 50 MeV, independent of the
magnetic field strength \teq{B}.

\section{INVERSE COMPTON IN THE GAMMA-RAY BAND}

Since significant questions surround the viability of the synchrotron
mechanism in explaining burst spectra, consideration of alternative
emission processes is warranted.  Here, the focus is on inverse Compton
scattering, another very efficient radiative mechanism.  For the
purposes of the discussion here, it comes in two relevant varieties:
scattering off a narrow-band soft photon distribution, and upscattering
of synchrotron photons (synchrotron self-Compton, or SSC) that provide
an inherently broad seed spectrum.  It shall be seen that the resulting
predictions for prompt emission spectra for these two cases are
manifestly different.  The inverse Compton (IC) mechanism possesses two
obvious advantages over synchrotron radiation: (i) high values of
\teq{B} can be avoided, since there is no requirement that the
synchrotron peak appear in the gamma-ray band, and (ii) the low energy
index can, in cases to be discerned in Sections~\ref{sec:ICformalism}
and~\ref{sec:alphalow_comp}, be as flat as \teq{\alpha =0}.  At the same
time, inverse Compton scenarios (particularly the SSC case) have the
disadvantage of requiring that synchrotron radiation or other components
be unobservable at optical and lower wavebands during the prompt and
early afterglow phases.  The parameter constraints thereby imposed are
discussed in Section~\ref{sec:constraint}.

The well-known criterion for IC to dominate synchrotron as an electron
cooling mechanism (i.e. also in total bolometric luminosity for a
spatially uniform medium) is simply that the energy density \teq{\us} of
the soft photons exceeds that of the ambient magnetic field, \teq{\ub
=B^2/8\pi} (e.g. Rybicki \& Lightman, 1979).  In general, the value of
\teq{\us} can vary enormously depending on the particular invocation for
soft photon generation.  For the specific case of a SSC model, \teq{\us}
couples to the field and can be determined by integrating
\teq{\erg\ndotsyn (\erg)} in Eq.~(\ref{eq:synch_emiss}) over \teq{\erg}
and multiplying by some fiducial photon escape timescale, \teq{\tesc
=R/c}, that reflects the physical size \teq{R} of the emission region.
For monoenergetic electrons of Lorentz factor \teq{\gammin\gg 1}, the
familiar result (e.g. see p.~168 of Rybicki \& Lightman, 1979) for the
synchrotron energy loss rate \teq{\Psyn} can be used to write
\teq{\us/\ub = n_e\tesc \Psyn /\ub = 4\sigt c\tesc n_e\gammin^2/3}.
Hence, the SSC emission can dominate the bolometric luminosity for
significant but not particularly large values of the product
\teq{n_e\gammin^2}.  It was precisely this property that led Gonzalez et
al. (2003) to argue that the intriguing hard gamma-ray component seen by
the EGRET TASC in GRB 941017 was not due to SSC emission, but perhaps
due to inverse Compton scattering seeded by an external soft photon
source.  The criterion for an SSC component to fall in the soft
gamma-ray band is \teq{\Gamma\gammin^4 B/B_{\rm cr}\sim 1}, which can
easily accommodate fields in the Gauss range for \teq{\Gamma\sim 300}
and \teq{\gammin\sim 100-10^3}.  A more thorough exploration of SSC
parameter spaces for burst prompt and afterglow emission can be found,
for example, in M\'esz\'aros, Rees, \& Papathanassiou (1994), Sari \&
Esin (2001) and Zhang \& M\'esz\'aros (2001).

\subsection{Inverse Compton Formalism}
 \label{sec:ICformalism}

The formalism for the inverse Compton (IC) emissivity in optically thin
environs is readily available in many astrophysical textbooks, such as
Rybicki \& Lightman (1979).  Since the description of IC polarization
effects is more subtle than for synchrotron radiation, being dependent
on anisotropies in the soft photon population, the developments here
are restricted to cases where photon polarizations are summed over.
The rate \teq{\ndotIC (\erg)} of emission, integrated over the angles
of the upscattered photons, is independent of the direction of momentum
of the seed photons.  Hence, no integration over \teq{\theta} is
required, contrasting Eq.~(\ref{eq:synch_emiss}) for synchrotron
radiation.  However, the rate is dependent on the soft photon energy
\teq{\es}, which must be integrated over, yielding the form (in units
of \teq{cm^{-3}\, sec^{-1}})
\begin{equation}
   \ndotIC (\erg )\; =\; \dover{3}{4}\sigt c
   \int_{0}^{\infty} d\es \, n_0 (\es )
   \int_{\gammin}^{\infty} d\gamma\, n_e(\gamma ,\;\theta )\;
   \dover{f(z)}{\gamma^2\es }\quad .
 \label{eq:IC_emiss}
\end{equation}
Here \teq{\sigt} is the Thomson cross section, \teq{n_0 (\es )}
is the energy distribution of soft photons, and 
\begin{equation}
  f(z)\; =\; \Bigl[ 2 z\log_ez +z+1-2 z^2 \Bigr]\;
  \Theta (z) \quad ,\quad
  z\; =\; \dover{\erg}{4\gamma^2\erg_{\rm s}}\;\; ,
 \label{eq:fdef}
\end{equation}
for a Heaviside step function \teq{\Theta (z)=1} when \teq{0\leq z\leq
1} and \teq{\Theta (z)=0} otherwise.  

Mirroring the treatment of synchrotron radiation, while numerical
evaluations of Eq.~(\ref{eq:IC_emiss}) were performed using the
electron distribution in Eq.~(\ref{eq:elec_dist}) to obtain
approximations to burst spectra in Section~\ref{sec:part_comp},
asymptotic spectral indices are readily attainable.  For monoenergetic
soft photons, the non-thermal contribution to the inverse Compton
spectrum (from the truncated power-law) is dominant whenever
\teq{\epsilon\gg \eta^{2+\delta}}, and assumes the following power-law
forms:
\begin{equation}
   \ndotIC (\erg )\Bigl\vert_{\rm NT} \propto
   \cases{ \Bigl(\dover{\erg}{\gammat^2 \es}\Bigr)^{0}\;
           \vphantom{\biggl(},&
              $\quad \es \ll\erg\ll \gammat^2 \es $,\cr
           \Bigl(\dover{\erg}{\gammat^2 \es}\Bigr)^{-(\delta +1)/2}\;
           \vphantom{\biggl(} ,&
              $\quad \gammat^2 \es \ll\erg $.\cr}
 \label{eq:ntIC_limits}
\end{equation}
The index at high energies, \teq{\erg\gg \gammat^2 \es}, is identical
to that for synchrotron radiation, since both mechanisms emit photons
with the same dependence on the electron Lorentz factor:
\teq{\erg\propto\gamma^2 \es}.  At low energies, the differential
spectrum is approximately independent of the upscattered energy of the
photon, due to the \teq{z\to 0} limit of \teq{f(z)} and the fact that
electrons near the \teq{\eta\gammat} cutoff dominate the low energy
photon production.  This index of \teq{\alpha =0} has been known since
the early days of application of inverse Compton to astrophysical
problems: see, for example, Figure~3 of the comprehensive treatment of
Jones (1968) that also illustrates that the Compton spectrum flattens
further to \teq{\erg^{+1}} in the domain \teq{\erg\lesssim\es} of
downscattering.

As in the case of synchrotron radiation, the thermal component, sampled
when \teq{\epsilon\ll \eta^{2+\delta}}, necessarily declines
exponentially at photon energies well above the thermal peak, but
possesses the same power-law dependence as in the non-thermal case for
low photon energies:
\begin{equation}
   \ndotIC (\erg )\Bigl\vert_{\rm TH} \propto \erg^{0},
   \quad \es \ll\erg\ll \gammat^2 \es \;\; .
 \label{eq:thIC_limits}
\end{equation}
This follows because the low energy photons are produced by electrons
mostly near the peak of the thermal distribution, again sampling
\teq{f(z)\approx} const.

\subsection{Synchrotron Self-Compton Formalism}
 \label{sec:SSCformalism}

From the perspective of spectral shapes at and below the gamma-ray peak,
it is important to explore the properties of synchrotron self-Compton
(SSC) radiation, to distinguish its character from IC emission that is
seeded by quasi-monoenergetic soft photons. Here the analysis will first
be confined to the case of a purely non-thermal electron distribution,
i.e. regimes \teq{\epsilon\gg \eta^{2+\delta}}, where the distribution
can be written in the form \teq{n_e(\gamma ,\;\theta ) =n_e\,
f_{\theta}\, (\delta -1)\, \gammin^{(\delta -1)}\, \gamma^{-\delta}} for
\teq{\gamma\geq\gammin} and is zero otherwise.  Hereafter, when making
the identification with the non-thermal portion of the distribution
function in Eq.~(\ref{eq:elec_dist}), \teq{\gammin =\eta\gammat}, and
one can set \teq{n_e\to n_{\theta}} and
\teq{f_{\theta}\to\epsilon\gammat\eta^{1-\delta }/(\delta -1)} to
establish a correspondence.

For such a truncated power-law, a compact analytic form for the SSC
spectrum was obtained, essentially reducing the quadruple integration to
one over a single variable.  Defining a fiducial photon escape
timescale, \teq{\tesc =R/c} for a one-zone emission region of size
\teq{R}, the average density of synchrotron photons that seed the SSC
emission can be written as \teq{n_0(\es ) =\tesc \ndotsyn (\es )}, so
that the form in Eq.~(\ref{eq:synchfinal}) can be directly inserted into
Eq.~(\ref{eq:IC_emiss}).  The analytic reduction is expounded in
Appendix A, with the result being
\begin{eqnarray}
   && \ndotssc (\erg )\Bigl\vert_{\rm NT} \; = \;
      \Ndotssc \; \dover{(\delta-1)^2}{(\delta+1)}
   \nonumber\\[-5.5pt]
 \label{eq:sscfinal}\\[-5.5pt]
   & \times & \Biggl\{
     \int_{0}^{\chi} dy \, K_{5/3}(y)
        P_{\delta} \Bigl(\dover{y}{\chi}\Bigr) +
     \int_{\chi}^{\infty} dy \, K_{5/3}(y)
        Q_{\delta} \Bigl(\dover{\chi}{y}\Bigr) \; \Biggr\}\;\; ,\nonumber
\end{eqnarray}
where 
\begin{equation}
   \Ndotssc \; =\; \; \dover{\fsc\sqrt{3}}{8\pi}\,
   \dover{c}{\lambar}\,\dover{\taut n_e}{\gammin^4}\, f_{\theta}^2\sin\theta
 \label{eq:Ndotsscdef}
\end{equation}
is the effective scale for the rate of the synchrotron-self-Compton process.
Here, \teq{\taut =n_e\sigt R} is the Thomson optical depth of the emission
region.  The emergent photon energy is encapsulated in a single 
dimensionless parameter
\begin{equation}
   \chi\; =\; \dover{\erg\, B_{\rm cr}}{6B_{\perp}\gammin^4}
   \quad , \quad B_{\perp}\; =\; B\sin\theta
 \label{eq:chidef}
\end{equation}
that denotes the energy scale of the peak for SSC emission, which
putatively could be in the gamma-ray band, and is the only spectral
structure in the emissivity.  The functions \teq{P_{\delta}(z)} and
\teq{Q_{\delta}(z)} are relatively compact and are given in
Eqs.~(\ref{eq:Pdef}) and~(\ref{eq:Qdef}) in Appendix A in terms of
elementary functions.  Eq.~(\ref{eq:sscfinal}) can be routinely
integrated numerically.

Asymptotic energy dependences of the SSC rate in Eq.~(\ref{eq:sscfinal})
are readily obtained.  In cases where \teq{\chi\gg 1}, the
\teq{Q_{\delta}} term never contributes to the integration since it is
exponentially suppressed by the behavior of the Bessel function. 
Accordingly, it is quickly deduced that the SSC spectrum approximately
obeys \teq{\ndotssc (\erg )\propto \erg^{-(\delta +1)/2}} at energies
\teq{\erg\gg 6\gammin^4 B_{\perp}/B_{\rm cr}}, as expected: a
convolution of an extended synchrotron spectrum with a truncated
power-law electron distribution with index \teq{(\delta +1)/2} is a
power-law of index \teq{(\delta +1)/2}.  There is a slight logarithmic
modulation as outlined in Eq.~(\ref{eq:SSCnt_highchi}).

The \teq{\chi\ll 1} limit can also be promptly obtained.  The
\teq{P_{\delta}} term samples only the \teq{y\ll 1} asymptotic limit of
the Bessel function, i.e.  the \teq{y^{-5/3}} form.  From this, it is
quickly verified that the \teq{y} integration is proportional to
\teq{\chi^{-2/3}}.  It is simple to show using the Eqs.~(\ref{eq:Pdef})
and~(\ref{eq:Qdef}) in Appendix A that the \teq{Q_{\delta}} term
possesses an identical dependence.  The net result is \teq{\ndotssc
\propto \erg^{-2/3}}, as embodied in Eq.~(\ref{eq:SSCnt_lowchi}), a
property that has been evident in various SSC models of bursts (e.g. see
Sari \& Esin, 2001).  These two limits can be summarized via
\begin{equation}
   \ndotssc (\erg )\Bigl\vert_{\rm NT} \propto
   \cases{ \Bigl(\dover{\erg\, B_{\rm cr}}{\gammat^4 B_{\perp}}\Bigr)^{-2/3}\;
           \vphantom{\biggl(},&
              $\erg\ll \gammat^4 \dover{B_{\perp}}{B_{\rm cr}} $,\cr
           \Bigl(\dover{\erg\, B_{\rm cr}}{\gammat^4 B_{\perp}} \Bigr)^{-(\delta +1)/2}
                \;  \log_e \dover{\erg\, B_{\rm cr}}{\gammat^4 B_{\perp}}
                \; \vphantom{\biggl(} ,&
              $\erg\gg \gammat^4 \dover{B_{\perp}}{B_{\rm cr}} $.\cr}
 \label{eq:ntSSC_limits}
\end{equation}
The low energy index images the synchrotron low energy tail, a result
that is not surprising since such a tail is steeper (and therefore its
constituent photons are more populous) than the tail resulting from IC
upscattering of monoenergetic soft photons in
Eq.~(\ref{eq:ntIC_limits}).  Accordingly, the bound \teq{\alpha < -2/3}
is placed on SSC spectra so that this mechanism cannot aid with the low
energy spectral index issues discussed in
Section~\ref{sec:alphalow_synch}.

\begin{figure*}[t]
\twofigureoutpdf{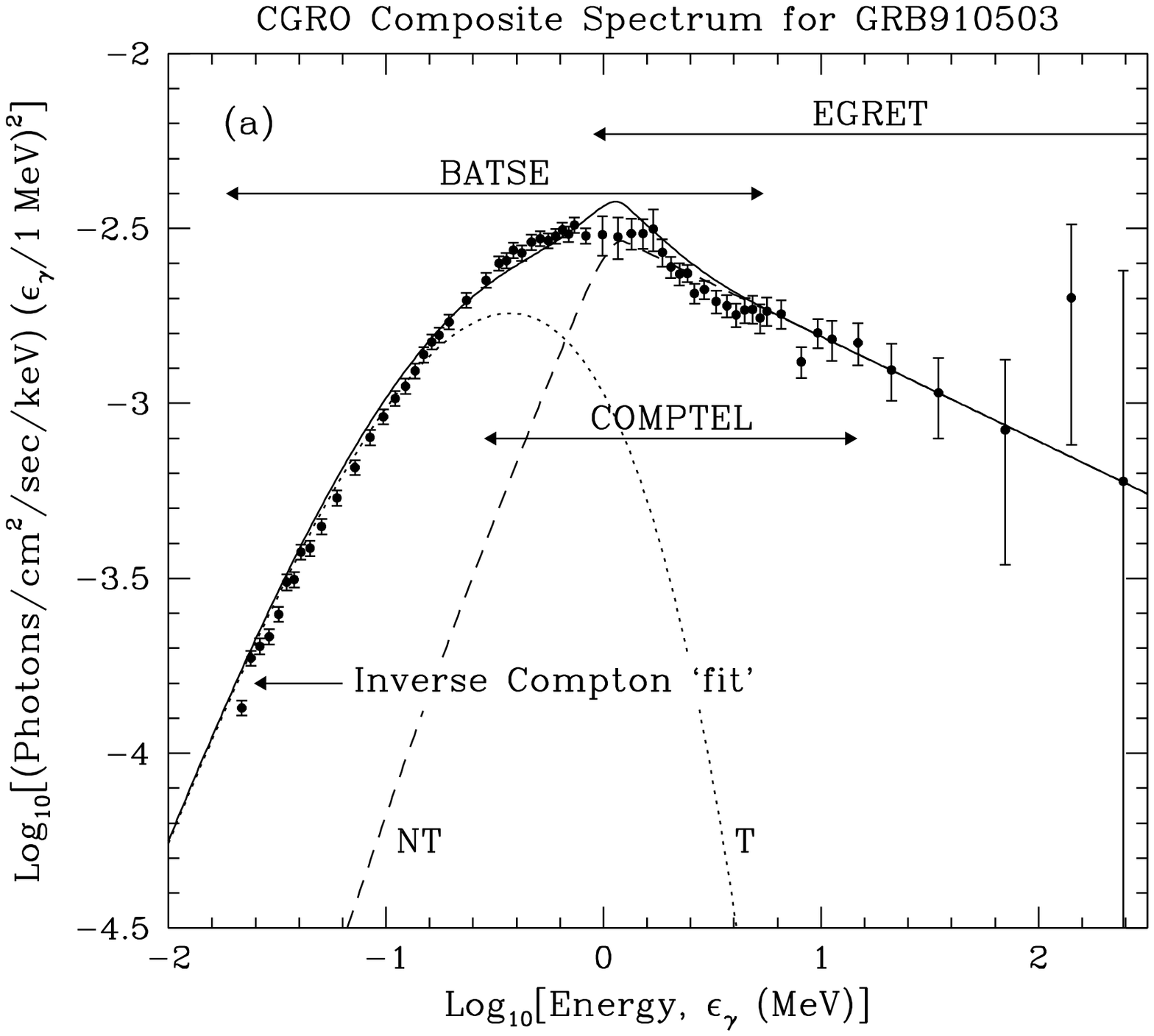}{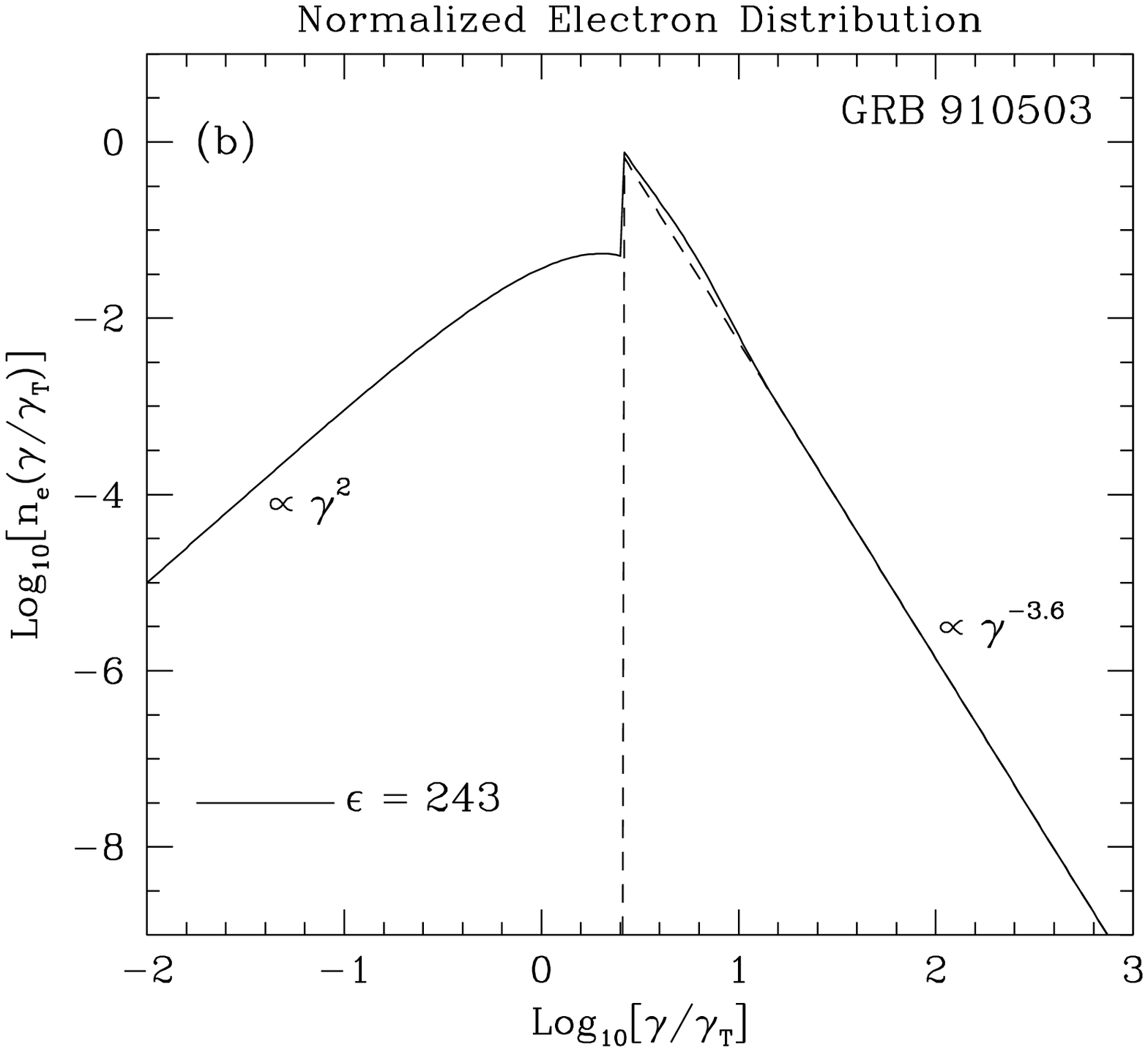}{ 
Photon spectra (a) from the pure inverse Compton model and the
corresponding normalized electron distribution (b) for the bright burst
GRB 910503 observed by the Compton Gamma-Ray Observatory.  The data
compilation and detecting instruments presented in (a) are as in
Fig~\ref{fig:synch_spec_dist}.  The solid curves correspond to the
particle distribution/resulting inverse Compton spectrum pair that
approximates the observed continuum very well [labelled Inverse Compton
'fit' in panel (a)]; model parameters included \teq{\eta=2.63} and
\teq{\epsilon\approx 243}.  The photon spectrum was generated using
monoenergetic soft photons of some energy \teq{\erg_0}.  In (a) the
thermal (labelled T) and non-thermal (labelled NT) components are
illustrated by the dotted and dashed curves, respectively.     In panel
(b), the total distribution is normalized to unit area as in
Fig~\ref{fig:synch_spec_dist}.  The dashed curve in (b) is the
non-thermal portion of the electron distribution, for which the index
was \teq{\delta=3.6}, and \teq{\Gamma\gammat^2 \es =1.32} arose from
adjustments along the energy axis.
 \label{fig:ic_spec_dist} }      
\end{figure*}

To facilitate the fitting of GRB spectra with SSC spectra, the case of
ultra-relativistic thermal particles must also be considered, for which
the distribution can be written in the form \teq{n_e(\gamma ,\;\theta )
=n_e\, f_{\theta}/ (2\gammat^3)\, \gamma^2\,\exp\{ -\gamma /\gammat\} }.
When making the identification with the thermal part of the distribution
function in Eq.~(\ref{eq:elec_dist}), one can set \teq{n_e\to
n_{\theta}} and \teq{f_{\theta}\to 2\gammat} to establish a
correspondence with correct normalization.  In Appendix B, an expedient
derivation of the emission spectrum is offered, making use of the result
already presented in Eq.~(\ref{eq:sscfinal}) for a truncated power-law
electron distribution.  Basically, the \teq{\delta\to\infty} limit is
taken to generate the spectrum for monoenergetic electrons, and this is
then convolved with a relativistic Maxwell-Boltzmann distribution. 
Setting
\begin{equation}
   \Ndotsscth \; =\; \; \dover{\fsc\sqrt{3}}{16\pi}\,
   \dover{c}{\lambar}\,\dover{\taut n_e}{\gammat^4}\, f_{\theta}^2\sin\theta
 \label{eq:Ndotsscthdef}
\end{equation}
as the effective scale for the rate of the synchrotron-self-Compton process
for the hot thermal particles, one arrives at
\begin{equation}
   \ndotssc (\erg )\Bigl\vert_{\rm TH} \; = \; \Ndotsscth \int_{0}^{\infty} dy \, K_{5/3}(y)\, 
   {\cal J}\Bigl( \Bigl[\dover{\chit}{y}\Bigr]^{1/4}\Bigr)\quad ,
 \label{eq:sscfinal_th_too}
\end{equation}
where the function \teq{{\cal J}(\kappa )} is defined in Eq.~(\ref{eq:calJdef}),
and is steeply declining as \teq{\kappa} increases.  Here,
\teq{\chit = \erg\, B_{\rm cr}/[6B_{\perp}\gammat^4]}.  Asymptotic forms
for the spectrum in Eq.~(\ref{eq:sscfinal_th_too}) are routinely derived,
being presented in Eq.~(\ref{eq:SSCTHlimits}).   As expected, an
\teq{\erg^{-2/3}} form arises at energies below the characteristic
energy \teq{ 6 B_{\perp}\gammat^4/B_{\rm cr}}, and the spectrum is a
weakly declining exponential far above this.

\subsection{Electron Distributions for Inverse Compton}
 \label{sec:part_comp}
 
Results for the spectral fitting of the GRB 910503 data for the case of
inverse Compton scattering are presented in
Figure~\ref{fig:ic_spec_dist}.  The fits were performed using the
formalism presented in Section~\ref{sec:ICformalism}, specifically with
numerical evaluations of Eq.~(\ref{eq:IC_emiss}). Again, the
multi-instrument, time-integrated CGRO data as published in Schaefer et
al. (1998) was used.  The figure displays both inverse Compton emission
spectra for the case of an isotropic, monoenergetic soft photon
distribution, and the inferred electron distribution that generates the
continuum. The solid curve depicts the case that would be required to
explain the photon spectrum, with the overall normalization again being
a free parameter; this curve was offset slightly from the data for
visual clarity.  The electron distributions were normalized to unit area
in \teq{\gamma/\gammat} space.

The non-thermal particle distribution clearly dominates the thermal
component in this ``fit,'' as is the case for synchrotron emission.  The
index \teq{\delta =3.6} was controlled by the EGRET data, and so is
identical to that in Fig.~\ref{fig:synch_spec_dist}. Adjustment of the
\teq{\eta} parameter to improve the appearance of the model
approximation around and above the spectral peak led to the assumed
value of \teq{\eta=2.63}, again accurate to around {15\%}.  The
positioning of the peak in energy produced the fit value
\teq{\Gamma\gammat^2 B_{\perp}/B_{\rm cr}=1.32}, where \teq{\Gamma} is
the Lorentz factor associated with the bulk motion of the GRB expansion.
The parameter \teq{\epsilon} describing the relative normalization of
the non-thermal and thermal components was found to be
\teq{\epsilon\approx 243} (\teq{\pm 10}\%), still a large value though
only a tenth of that obtained for the synchrotron fit.

The conclusion is essentially identical to that in
Section~\ref{sec:part_synch}: the observational data can only be
accommodated by an electron distribution that has a preponderance of
non-thermal electrons as Fig.~(\ref{fig:ic_spec_dist}b).  Such a
determination was also obtained for BATSE bursts GRB 940619 (trigger
number 3035\_6) and GRB 940817 (trigger number 3128\_5) from the Preece
et al. (2000) catalog. Moreover, the fits in
Figs.~\ref{fig:synch_spec_dist} and~\ref{fig:ic_spec_dist} are of
comparable quality, at least below 1 MeV and above 5 MeV; i.e. there is
little potential in the CGRO data to discriminate between the two
emission mechanisms.  There is a small bump/excess in the theoretical
spectrum in Fig.~\ref{fig:ic_spec_dist} just above 1 MeV that is a
consequence of the sharper inverse Compton scattering kernel; its
prominence can be muted somewhat by slightly reducing both
\teq{\epsilon} and the index \teq{\delta}.  Such a liberty may not be
afforded in the GLAST era when burst spectral indices will be better
constrained because of its broad energy range, sampled by the
Gamma-Ray Burst Monitor (GBM) in the BATSE band and the Large Area
Telescope (LAT) in the EGRET band.

The last point to note is that the inverse Compton fit is tolerated here
because the low energy asymptotic index is possibly not fully realized
in the BATSE window.  The data possess a slight curvature that may be
consistent with a continued flattening of the spectrum below 30 keV, as
would be expected for the inverse Compton mechanism.  Yet the data match
the synchrotron emissivity also, indicating that BATSE cannot
discriminate between these two mechanisms on the basis of the soft
gamma-ray portion of the bandwidth.  This impasse should be addressed when
Swift is launched in the coming year.  Swift possesses the sensitivity
in the critical 1 -100 keV band to probe the spectrum below
the \teq{\nu F_{\nu}} peak and afford diagnostics on the emission
mechanism.  Furthermore, the spectral coverage afforded by the GBM 
and the LAT on GLAST will also provide powerful observational 
constraints.

\subsection{Low Energy Spectral Indices}
 \label{sec:alphalow_comp}

The low energy asymptotic forms in Eqs.~(\ref{eq:ntIC_limits})
and~(\ref{eq:thIC_limits}) raise the possibility that
inverse Compton emission can eliminate the so-called ``line of death''
problem that is pervasive in synchrotron models.  The IC mechanism can
accommodate low energy asymptotic indices as flat as \teq{\alpha =0},
with steeper indices easily being generated by appropriately
distributed electrons.  The domain of compatibility is indicated in
Fig.~\ref{fig:BATSE_low_index}, where around 98\% of BATSE burst
spectra in the spectroscopy catalogue of Preece et al. (2000) can be
accounted for by the inverse Compton process.  This is an enticing
property, but is not a definitive statement of the operation of this
mechanism in GRB prompt emission regions.  It must be emphasized that
the establishment of an \teq{\alpha\geq 0} domain is contingent upon
the soft photons that seed the upscattering assuming a narrow energy
(i.e. quasi-monoenergetic) distribution.  This caveat is exemplified by
the SSC value of \teq{\alpha} being identical to that for synchrotron
radiation due to the breadth of the synchrotron seed spectrum.  Hence
invoking a synchrotron self-Compton model only dramatically impacts the
operating range of \teq{B} and not the spectral shape in the 
very soft gamma-rays.

Given the discussion in Section~\ref{sec:anis_synch} pertaining to
synchrotron radiation, it is natural to ask whether anisotropies in
either the soft photon or electron populations can dramatically alter
the spectrum.  The inverse Compton photon production rate in
Eq.~(\ref{eq:IC_emiss}) implicitly assumes isotropy of the soft photons
constituting \teq{n_0(\es )}. Furthermore, it assumes that the angular
distribution of electrons is not seriously deficient along the line of
sight to the observer; i.e. that the viewing perspective is not towards
the side of an electron jet.  Calculating IC spectra for anisotropic
distributions is quite involved, and moreover sensitive to the choice
of geometry.  Yet spectral behavior depends mostly on the interplay
between such geometry and the kinematics of IC scattering.

If the electron distribution is highly collimated and the observer's
direction lies outside the cone of jet collimation, then the kinematics
of scattering requires a smaller angle \teq{\theta_{\rm scatt}} of
scattering in the electron rest frame to send the photon towards the
observer.  This small \teq{\theta_{\rm scatt}} yields a lower final
photon energy in the observer's frame, so that the emitted inverse
Compton spectrum is softer outside the jet core.  The spectrum then
flattens at lower energies (e.g. see Dermer \& Schlickeiser 1993; Baring
1994 for blazar contexts) due to the kinematic depletion of photons in
the observer's direction, and can be even slightly flatter than
\teq{\erg^{0}}.   A problem with this scenario immediately arises: the
energetics are dominated by emission in the jet cone so that assuming
``off-axis'' IC emission places more severe constraints on the GRB
radiative efficiency.  Note that fluctuations in the jet orientation or
structure would lead to the emergent inverse Compton spectra being
highly variable both in flux (observed in bursts and therefore not a
problem, in principle) and in spectral shape --- any structure such as
the \teq{\nu F_{\nu}} peak would be expected to vary in time rapidly in
energy due to the scattering kinematics.  While modest fluctuations (as
opposed to hard-to-soft evolution) are present in gamma-ray burst prompt
emission, detailed modeling would be required to discern whether they
could match those produced by anisotropic scattering scenarios.

\begin{figure*}[t]
\twofigureoutpdf{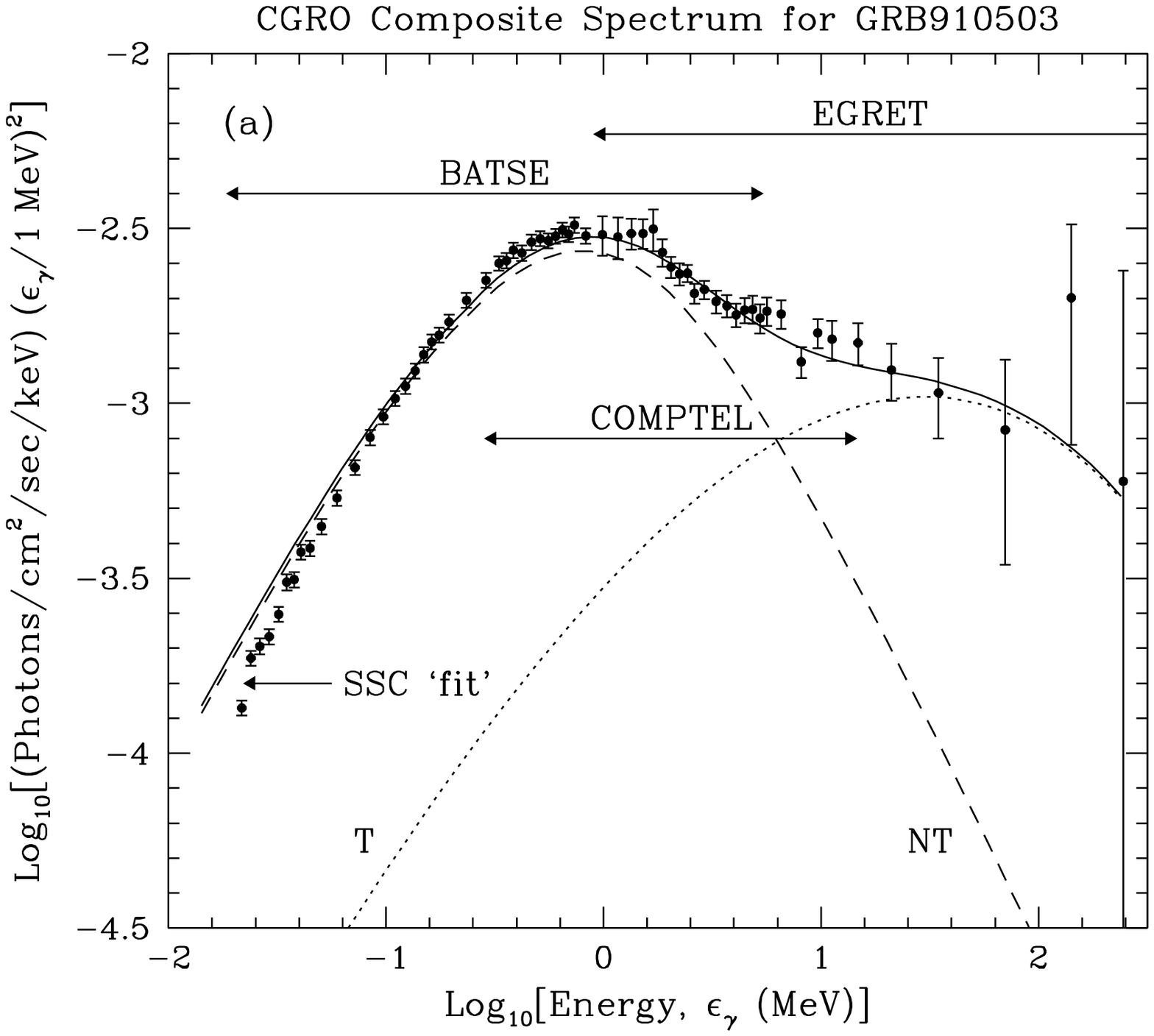}{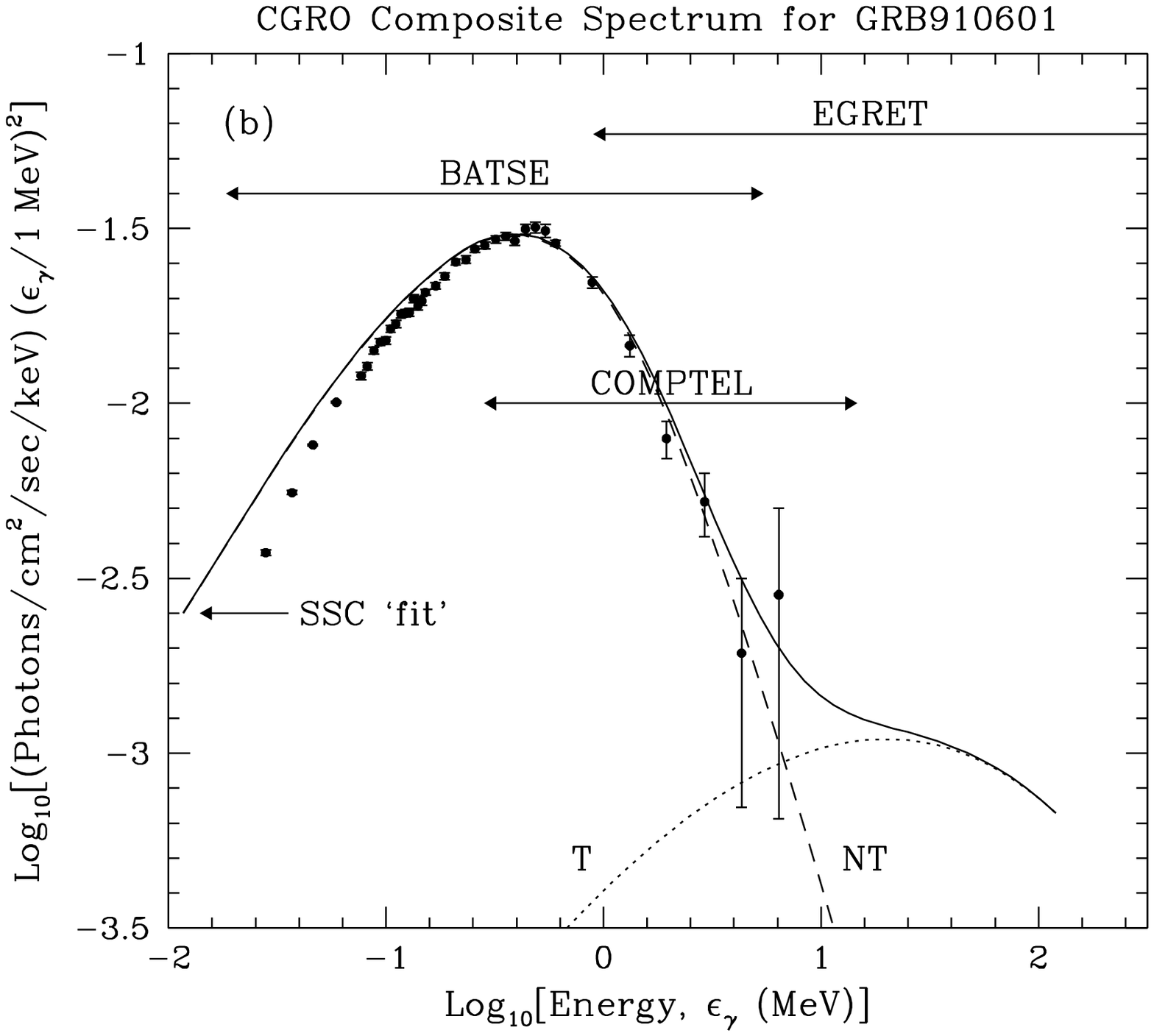}{ 
Photon spectra from the synchrotron self-Compton model for the bright
bursts (a) GRB 910503 and (b) GRB 910601.  The data compilations from
the Compton Gamma-Ray Observatory (the operating energy bands of the
three detecting instruments, BATSE, COMPTEL and EGRET, are again as
indicated) are from Schaefer et al. (1998).  The solid curves correspond
to the resulting synchroton self-Compton spectrum that best approximates
the observed continuum [labelled SSC 'fit']. Model parameters included
\teq{\delta=6.0}, \teq{\eta=2.0} and \teq{\epsilon\approx 15} for GRB
910503, and \teq{\delta=8.0}, \teq{\eta=2.0} and \teq{\epsilon\approx
200} for GRB 910601.  In both panels the thermal (labelled T) and
non-thermal (labelled NT) components are illustrated by the dotted and
dashed curves, respectively.     The small thermal contributions in both
cases render the values of \teq{\epsilon} quite uncertain, so that the
depicted values should be regarded as approximate lower bounds to
\teq{\epsilon}.  Adjustments along the energy axis yielded
\teq{\Gamma\gammat^4 B_{\perp}/B_{\rm cr}=5.0\times 10^{-3}} for GRB
910503 and \teq{\Gamma\gammat^4 B_{\perp}/B_{\rm cr}=3.3\times 10^{-3}}
for GRB 910601, both implying sub-Gauss fields for typical values of
\teq{\Gamma\sim\gamma\sim 300}.
 \label{fig:ssc_spec_dist} }      
\end{figure*}

Any anisotropy in the soft photon population is not sampled until they
become more collimated than the electrons; only then does the emissivity
change appreciably from Eq.~(\ref{eq:IC_emiss}).  Then, electrons within
the photon ``cone'' of beaming experience an almost full range of
incident photon angles in their rest frames, so that they contribute an
inverse Compton spectrum similar to that for isotropic soft photons. 
However, the majority of electrons lie outside this cone, and for these,
collisions with photons are predominantly head-on, leading to a
depletion of the softer photons from the overall inverse Compton
spectrum just as with electron collimation.  Hence, one expects that
such soft photon anisotropies could in principal also flatten the
spectrum below the \teq{\nu F_{\nu}} peak.

\subsection{Electron Distributions for SSC}
 \label{sec:part_SSC}
 
To provide a meaningful comparison with the preceding mechanisms, GRB
910503 is again used as a focal point.  Yet, so as to provide a
variation, here a case study for another EGRET burst, GRB 910601 is also
presented. Results for the spectral fitting for synchrotron self-Compton
scattering are presented in Figure~\ref{fig:ssc_spec_dist}.  The
``fits'' were performed using the formalism presented in
Section~\ref{sec:SSCformalism}, specifically with numerical evaluations
of Eqs.~(\ref{eq:sscfinal}) and~(\ref{eq:sscfinal_th_too}). The figure
displays the SSC emission spectra for the case of an isotropic electron
distribution, for each burst. The solid curve depicts the case that
would most closely explain the photon spectrum, with the overall
normalization again being a free parameter. Due to the strong dependence
of the characteristic energy on electron Lorentz factor, the values of
\teq{\Gamma\gammat^4 B_{\perp}/B_{\rm cr}} required to move the SSC peak
into the gamma-ray band lead to inferences of sub-Gauss fields if
\teq{\gammat\sim\Gamma\gtrsim 300}.

Consider first GRB 910503.  The most obvious features are that the
``fit'' is more difficult than for the emission mechanisms previously
considered, with poor results for the lower energies much below the
peak, and that the thermal contribution must be dramatically suppressed.
The cause is easily identified: the SSC spectrum is intrinsically much
broader than either the IC or synchrotron spectra.  This is due to the
characteristic energy for SSC embodied in the dimensionless parameter
\teq{\chi} (see Equation~[\ref{eq:chidef}]) scaling as \teq{\gammin^4}
(or \teq{\chit\propto\gammat^4} for the thermal case).  Consequently,
small changes in the electron \teq{\gamma} near the peak of the
Maxwell-Boltzmann distribution incur large variations in the SSC photon
energy so that the spectral peak is broad.  This effect is manifested in
the non-thermal SSC emissivity in Eq.~(\ref{eq:sscfinal}), which leads
to the excesses evident around 30 keV in the ``fits'' of
Figure~\ref{fig:ssc_spec_dist}, and more profoundly for the thermal SSC
contribution, which is extremely broad. From Eq.~(\ref{eq:SSCTHlimits})
a notable conclusion can be drawn: that the thermal SSC spectrum in the
depicted \teq{\nu F_{\nu}} representation only starts to fall above
around 30 MeV, and then only slowly, i.e. it is blueward of the peak
from the non-thermal SSC contribution.  This drives the requisite
dramatic suppression of the normalization of a thermal component to the
underlying electron distribution, leading to the same issues (but in
this case more strikingly emphasized) for acceleration from thermal
energies that were discussed above.  Note that for values of
\teq{\epsilon >25}, a comparably satisfactory SSC fit can be obtained
with \teq{\delta\sim 5}. Observe that for such a large \teq{\delta}, the
tail of the thermal Maxwellian emerges above that of the non-thermal
distribution for a small range of Lorentz factors \teq{\gamma\gtrsim
3\gammat}.

The situation for GRB 910601 is even more constrained, due to the
markedly steeper gamma-ray spectrum.  The thermal electrons must be even
less populous to accommodate the Comptel and EGRET data, and the
non-thermal electrons must possess a very steep index \teq{\delta =8.0},
considerably steeper than the \teq{\delta=5.5} required for a
synchrotron fit.  In fact, both the synchrotron and IC models had much
greater success fitting the GRB 910601 data, though are not depicted. 
Steepening the index to \teq{\delta\gtrsim 10} produces only marginal
improvement in the fit. The inevitable conclusion of this analysis is
that the breadth of the SSC emissivity for even monoenergetic electrons
(derived in Equation~(\ref{eq:sscfinal_mono}), and closely approximated
by the \teq{\delta =8.0} case in Figure~\ref{fig:ssc_spec_dist}) is
large enough to provide significant or severe problems for the SSC
interpretation of prompt emission from gamma-ray bursts. This contention
is not based solely on the GRB 910503 and in particular the GRB 910601
data; the similarity of the spectra of GRB 910814 (Schaefer et al. 1998)
and GRB 990123 (Briggs et al. 1999) to that of GRB 910601 strongly
suggests that a spectroscopic incompatibility of the SSC mechanism for
prompt emission may be commonplace, specifically for differential burst
spectra steeper than around \teq{E^{-2.5}} above the \teq{\nu F_{\nu}}
peak.  Bursts such as the Superbowl one (GRB 930131) with high energy
index \teq{\beta\approx -2} may be much more compatible with an SSC
scenario.

While this assessment may achieve closure in the Swift/GLAST era, a
caveat is possibly provided if the synchrotron continuum is strongly
self-absorbed.  This is the scenario envisaged by Panaitescu \& Meszaros
(2000) and Liang, Boettcher \& Kocevski (2003), with the latter group
suggesting that SSC with moderate optical depths (\teq{\taut\sim
0.1}--\teq{0.4}) could explain X-ray excesses seen in some burst
spectra.  With such strong absorption, the supply of soft photons for
inverse Compton upscattering is more sharply peaked than even the
synchrotron continuum from electrons possessing a truncated power-law
distribution, thereby approximately mimicking the pure inverse Compton
spectral shapes explored in Section~\ref{sec:part_comp}. The requirement
that self-absorption be effective in narrowing the synchrotron
self-Compton peak is that the turnover energy \teq{\erg_t} exceeds that
of the synchrotron peak, which is below the X-ray band.  Using
Eq.~(\ref{eq:selfabs_energy}), one quickly arrives at the requisite
condition:
\begin{equation}
   \eta\gammat\;\lesssim\; \omega (\delta )\; \dover{\taut}{\fsc}\, 
   \dover{\Gamma}{\erg_{\sevenrm peak}}\quad ,
 \label{eq:SSC_selfabs}
\end{equation}
for
\begin{equation}
   \omega (\delta )\; =\; 2^{-(\delta+2)/2} \, \sqrt{3}\, (\delta -1)\;
   \Gamma \biggl( \dover{\delta}{4}+\dover{1}{6} \biggr)\, 
   \Gamma \biggl( \dover{\delta}{4}+\dover{11}{6} \biggr)\;\; . 
 \label{eq:omega_def}
\end{equation}
Here \teq{\erg_{\sevenrm peak}\lesssim 1} is the energy of the \teq{\nu
F_{\nu}} peak in the prompt emission.  For \teq{2\lesssim \delta\lesssim
5}, one deduces that \teq{\omega (\delta )/\fsc \sim 110}--\teq{160}, so
that optical depths \teq{\taut <1} can be tolerated for typical
\teq{\Gamma \sim 300} without pushing \teq{\gammat} too low, i.e. into
the mildly-relativistic regime.  Detailed computations of self-absorbed
SSC spectra are deferred to future work.

\subsection{Broadband Observational Constraints}
 \label{sec:constraint}

In addition to the soft gamma-ray spectral shape, a probing diagnostic
on the emission mechanisms is whether their extrapolation down to the
optical band can be accommodated by optical monitoring data for bursts. 
Such broadband constraints turn out not to be profoundly limiting for
synchrotron or pure inverse Compton emission.  However, they do impact
SSC models, since these can potentially have a prominent synchrotron
component in the optical that might violate upper bounds or, in the case
of GRB 990123, prompt optical detections.  It is straightforward to
assess how optical band data can constrain SSC model parameters using
the formalism derived here.

First, the focus is on observational data.  GRB 990123 (with redshift
\teq{z=1.6}) is to date the only burst with a prompt optical detection
(Akerlof et al. 1999), peaking at around magnitude 8.8 some 47 seconds
after BATSE trigger and contemporaneous with the BATSE signal.  These
prompt observations were made by the Robotic Optical Transient Search
Experiment-I (ROTSE-I). The recent bright burst GRB 030329 (redshift
\teq{z=0.168}) was only detected optically by ROTSE-III in the afterglow
epoch at visual magnitude 12.5 over an hour after the burst trigger on
HETE-2 (Smith et al. 2003).  Kehoe et al. (2001) report upper bounds of
around magnitude 13--15 to half a dozen bursts, in two cases starting
just 15-20 seconds after trigger.  These low limiting magnitudes can
potentially be severely constraining on models with a high optical to
gamma-ray flux expectation, though contemporaneous multiwavelength
detections are requisite for unambiguous diagnostics.

Taking magnitude 9 as a benchmark optical signal for the purposes of
exploring broadband constraints, a number that must be considered modulo
the source distance and intrinsic luminosity, this corresponds to a
differential photon flux of the order of \teq{10^6} photons sec$^{-1}$
cm$^{-2}$ keV$^{-1}$ at around 2eV.  This can be compared with the BATSE
spectrum of GRB 910503, a burst brighter than most of those in the
sample of Kehoe et al. (2001), which has a time-integrated differential
flux of \teq{0.1} photons sec$^{-1}$ cm$^{-2}$ keV$^{-1}$ at around 100
keV.  The interpolation between such an optical magnitude and gamma-ray
flux is a roughly \teq{\erg^{-1.5}} spectrum. Immediately, one infers
that synchrotron and inverse Compton models for the gamma-rays can
accommodate such an optical magnitude; only limiting magnitudes of
around 19 or higher can provide conflicts for these mechanisms.  Note
that a similar conclusion would apply for pitch angle synchrotron models, 
which possess very flat low energy asymptotic spectra like
inverse Compton scenarios.

The situation for SSC broadband spectra is very different.  The ratio of
the energy of the SSC peak to that of the synchrotron peak is simply of
the order \teq{4\gammin^2}, which would set the synchrotron peak in the
infra-red for Lorentz factors of \teq{\gammin\sim 10^3}, independent of
the bulk flow \teq{\Gamma}.  Lower \teq{\gammin} can move the
synchrotron peak into the optical and even the UV.  The ratio of
emissivities at the synchrotron and SSC peaks is quickly determined 
from Equations~(\ref{eq:Ndotsyn_def}) and~(\ref{eq:Ndotsscdef}) to
be \teq{\Ndotsyn /\Ndotssc\sim 8\gammin^2/(\taut f_{\theta})} in the
notations of this paper, i.e. with
\teq{f_{\theta}\to\epsilon\gammat\eta^{1-\delta }/(\delta -1)}. Hence,
for \teq{\taut f_{\theta} \sim 1}, the peaks of the two components would
lie roughly on an \teq{\erg^{-1}} spectrum.  Moreover, opting for
\teq{\gammin\sim 300} to position a synchrotron peak in the optical
window, normalizing to the BATSE spectrum of GRB 910503 would yield a
count rate an order of magnitude below the magnitude 9 benchmark posited
above.  At face value, this would render the SSC scenario compatible
with limiting magnitudes of around 11 for this source.  Yet
Comptonization of the spectrum is not evident in the gamma-rays, so it
is inferred that \teq{\taut\ll 1}, thereby increasing the ratio
\teq{\Ndotsyn /\Ndotssc} and lowering the permissible limiting magnitude
for a viable SSC model. Accordingly, for the only source with a prompt
optical detection, GRB 990123, for the SSC model not to overestimate the
observed optical flux, the synchrotron peak has to move somewhat out of 
the optical, though not much.  This amounts to a modest constraint in 
\teq{(\gammin ,\, \taut ,\, \delta )} phase space that can be quickly deduced,
though at this juncture is not particularly enlightening; such an exercise will
be more timely once Swift is providing a number of simultaneous hard
X-ray and optical detections of bursts.

\section{CONCLUSION}
 \label{sec:conclusion}

The interpretative modeling presented here provides a strong indication
that the synchrotron and inverse Compton processes may provide a more
viable explanation of prompt burst spectra than the SSC mechanism. The
spectroscopic bias against SSC is due to its inherently broad character
around the \teq{\nu F_{\nu}} peak, and is only marginally constrained by
broad-band considerations, i.e. limiting magnitudes in the optical band.
However, strong self-absorption of the synchrotron component can provide
an escape clause for SSC scenarios, though this imposes significant
fine-tuning of parameter space. Inverse Compton provides an attractive
possibility for the vast majority of bursts, as does pitch angle
synchrotron emission, though quasi-isotropic synchrotron radiation may
only be reconciled with around 2/3 of bursts at the lowest energies in
the BATSE window. Discrimination between these processes may prove
possible after Swift launches and in the GLAST era, though polarimetric
measurements would enhance the diagnostic capability of observations. 
In analyzing SSC spectral models, attractively compact analytic forms
for the synchrotron self-Compton emissivity from thermal and truncated
power-law electron distributions were developed; these will prove useful
for applications of SSC spectral models to a variety of astrophysical
environments.

A distinctive conclusion of this work is that agreeable ``fits'' with
either synchrotron or inverse Compton emission scenarios are only
attainable if the underlying electron distribution is profoundly
dominated by a non-thermal component.  This result is in
contradistinction with standard understanding of acceleration mechanisms
such as the diffusive (Fermi) process at shocks, where the non-thermal
particles are drawn probabilistically from the shocked thermal
population, and therefore are subdominant in number.  This constraint
has profound implications for any proposed GRB model, providing a
challenge for theorists in an era when computational resources are
permitting new avenues for exploration using intensive numerical
simulations of particle acceleration at shocks.

\acknowledgments 
This work was supported by the NSF Extragalactic Astrophysics Program.
Matthew Braby acknowledges the support of a Shell Foundation Scholarship
towards this research during his undergraduate degree at Rice University.
We thank the referee, Rob Preece, for insightful and constructive suggestions
for polishing the paper, and Nicole Lloyd-Ronning, Jerry Fishman and 
Demos Kazanas for reading through the manuscript and providing helpful comments.
We also thank Brad Schaefer for supplying the GRB 910503 compilation data
prior to its publication in 1998.


\appendix
\section{Reduction of the Synchrotron Self-Compton Rate for 
Non-Thermal Electrons}
 \label{sec:appendixA}

In this Appendix, developments leading to the form in
Eq.~(\ref{eq:sscfinal}) for the rate of synchrotron self-Compton
emission from a truncated power-law distribution of electrons are
expounded.  The starting point is the insertion of \teq{\tesc} times
the synchrotron rate in Eq.~(\ref{eq:synchfinal}), serving as the soft
photon density, directly into the expression in Eq.~(\ref{eq:IC_emiss})
for the inverse Compton emissivity.  This presents a triple integral
that can be reduced to a compact single integration by appropriate
manipulations.  The Lorentz factor \teq{\gamma} of the electron that
upscatters the synchrotron photons proves to be a less convenient
integration variable than \teq{z=\erg/(4\gamma^2\es )}, which is defined 
in Eq.~(\ref{eq:fdef}).  Likewise, to proves expedient to use
\begin{equation}
  q\; =\; \dover{2\es\, B_{\rm cr}}{3B\, y\gammin^2\sin\theta} \;\; ,
 \label{eq:q_def}
\end{equation}
as an integration variable, rather than the synchrotron photon energy
\teq{\es}.  Performing these two changes of variables leads quickly to 
an SSC emissivity of
\begin{equation}
   \ndotssc (\erg )\Bigl\vert_{\rm NT}\; =\; \Ndotssc
   \; \dover{(\delta-1)^2}{(\delta+1)}
   \int_{0}^{\infty} dy \, K_{5/3}(y)\, \int_{0}^{\infty} \dover{dq}{q}\, 
   \biggl( \dover{qy}{\chi\xi} \biggr)^{(\delta +1)/2}\;
   g_{\delta}(\zmax)\;\; , 
 \label{eq:ssc_take3}
\end{equation}
where \teq{\Ndotssc} is the SSC rate scale factor given in 
Eq.~(\ref{eq:Ndotsscdef}), and \teq{\chi} is the dimensionless SSC energy
variable defined in Eq.~(\ref{eq:chidef}).  Here, \teq{\xi} and the 
maximum value \teq{\zmax} of \teq{z} can be written
\begin{equation}
   \xi\; =\;\max \{ 1,\, q\} \quad ,\quad
   \zmax\; =\; \min \Biggl\{ 1 ,\; \dover{\chi}{yq} \Biggr\}\;\; .
 \label{eq:xi_zmaxdef}
\end{equation}
The \teq{z}-integration over the inverse Compton scattering kernel
leads to a definition of \teq{g_{\delta}}:
\begin{eqnarray}
   g_{\delta}(z) &=& \int_0^z f(q) q^{(\delta -1)/2}\, dq
   \; \equiv\; z^{(\delta +1)/2}\, G_{\delta}(z)\quad ,\nonumber\\[-5.5pt]
 \label{eq:gzdef}\\[-5.5pt]
   G_{\delta}(z) & =& \dover{2z}{\delta+3}\;
   \biggl( 2\log_ez + \dover{\delta-1}{\delta+3} \biggr)
   + \dover{2}{\delta+1} - \dover{4z^2}{\delta+5}\quad ,\nonumber
\end{eqnarray}
The \teq{q} integration can be reduced by noting that the
definitions of \teq{\xi} and \teq{\zmax} establish two critical
values of \teq{q}, namely \teq{q_1 = \chi/y} and \teq{q_2=1}.
which delineate distinct dependences of the integrand on \teq{q}.
The relative ordering of \teq{q_1} and \teq{q_2} depends on
the value of \teq{y} assumed:  \teq{y >\chi \Rightarrow q_2 >q_1} and
\teq{y <\chi \Rightarrow q_2 <q_1}.  This naturally divides
the \teq{y} integration into two ranges, \teq{[0,\,\chi ]}, and
\teq{[\chi ,\,\infty )}.   The developments are straightforward,
and are facilitated using an \teq{h_{\delta}} function defined via
\begin{eqnarray}
   h_{\delta}(z) &=& \int_0^z \dover{g_{\delta}(q)}{q}\, dq
   \; \equiv\; z^{(\delta +1)/2}\, H_{\delta}(z)\quad ,\nonumber\\[-5.5pt]
 \label{eq:hzdef}\\[-5.5pt]
   H_{\delta}(z) & =& \dover{4z}{(\delta+3)^2}\;
   \biggl( 2\log_ez + \dover{\delta-5}{\delta+3} \biggr)
   + \dover{4}{(\delta+1)^2} - \dover{8z^2}{(\delta+5)^2}\quad .\nonumber
\end{eqnarray}
The \teq{y >\chi} case requires an integration by parts in one
contributing \teq{q} interval, and one quickly arrives at the compact
form 
\begin{equation}
   \ndotssc (\erg )\Bigl\vert_{\rm NT} \; = \; \Ndotssc \; \dover{(\delta-1)^2}{(\delta+1)}
   \Biggl\{
     \int_{0}^{\chi} dy \, K_{5/3}(y) P_{\delta} \Bigl(\dover{y}{\chi}\Bigr) +
    \int_{\chi}^{\infty} dy \, K_{5/3}(y) Q_{\delta} \Bigl(\dover{\chi}{y}\Bigr)
 \; \Biggr\}\;\; ,
 \label{eq:sscfinal_app}
\end{equation}
i.e. Eq.~(\ref{eq:sscfinal}), using the following functional forms
(with \teq{G_{\delta}} and \teq{H_{\delta}} defined as just above):
\begin{equation}
    P_{\delta}(z) \; =\; z^{(\delta+1)/2}\,
          \Biggl\{\, \biggl[ \dover{2}{\delta +1}
           - \log_e z \biggr]\, G_{\delta}(1) + H_{\delta}(1)\,\Biggr\} \;\; ,
 \label{eq:Pdef}
\end{equation}
for \teq{z\to 1/q_1 = y/\chi}, and for \teq{z\to q_1=\chi /y}
\begin{eqnarray}
    Q_{\delta}(z)  &=& \Biggl\{\, \dover{2}{\delta +1}\biggl[ G_{\delta}(z)
          + (z+2)(z-1) - (1+2z)\,\log_e z \biggr] + H_{\delta}(z)\,\Biggr\} \quad ,\nonumber\\[-5.5pt]
 \label{eq:Qdef}\\[-5.5pt]
       &=& - 2z\,\dover{ \delta +1}{(\delta +3)^2}\biggl[ 2 \,
           \log_ez - \dover{\delta +11}{\delta +3} \biggr]
      - 4 \dover{\delta -1}{(\delta +1)^2} -2 \dover{\log_ez}{\delta +1}
       + 2z^2\, \dover{\delta +1}{(\delta +5)^2}\quad .\nonumber
\end{eqnarray}
Observe that as \teq{z\to 1}, \teq{P_{\delta} \to Q_{\delta}}, as
demanded by continuity of the spectrum.  


Asymptotic dependences of the SSC rate on the upscattered photon energy, 
encapsulated in the proportionalities of Eq.~(\ref{eq:ntSSC_limits}), can be
simply derived.  In the limit \teq{\chi\ll 1}, both the \teq{P_{\delta}} and 
\teq{Q_{\delta}} terms contribute to Eq.~(\ref{eq:sscfinal_app}), with the integrals
being dominated by the \teq{y\ll 1} domain.  The Bessel function can be
replaced by its power-law asymptotic form \teq{K_{5/3}(z)\approx 2^{2/3}
\Gamma (5/3)\, y^{-5/3}}, resulting in elementary integrals.  For \teq{\delta >1/3}
the integration converges to
\begin{equation}
   \ndotssc (\erg )\Bigl\vert_{\rm NT} \; = \; 
   \dover{567}{50}\, 2^{2/3}\Gamma (2/3) \; 
   \dover{(\delta-1)^2}{(3\delta-1)^2} \; 
   \dover{ \Ndotssc}{\chi^{2/3}}\quad ,\quad \chi\,\ll\, 1\;\; ,
 \label{eq:SSCnt_lowchi}
\end{equation}
The \teq{\chi\gg 1} case is less simple, though the integration over \teq{Q_{\delta}} 
does not contribute significantly since it is exponentially suppressed by the
dependence of the Bessel function.  The \teq{P_{\delta}} integration can be developed
using identity 6.561.16 of Gradshteyn and Ryzhik (1980), which can be specialized
to the case appropriate for this analysis:
\begin{equation}
   I(\mu )\; \equiv\; \int_0^{\infty} x^{\mu}\, K_{5/3}(x)\, dx
   \; =\; 2^{\mu -1}\;
   \Gamma \biggl( \dover{\mu}{2} + \dover{4}{3}\biggr)\,
   \Gamma \biggl( \dover{\mu}{2} - \dover{1}{3}\biggr)\quad , \quad \mu\, >\, \dover{2}{3}\;\; .
 \label{eq:Besselint_ident}
\end{equation}
The derivative of this with respect to \teq{\mu} can also be used to
expedite the developments, and can be written in the form 
\begin{equation}
   \dover{dI}{d\mu}
   \; \equiv\; \int_0^{\infty} x^{\mu}\, \log_ex\, K_{5/3}(x)\, dx
   \; =\; \dover{1}{2}\, I(\mu )\; \Biggl\{
   \psi \biggl( \dover{\mu}{2} + \dover{4}{3}\biggr) +
   \psi \biggl( \dover{\mu}{2} - \dover{1}{3}\biggr) + 2\log_e2 \Biggr\}\quad ,
 \label{eq:Besselint_ident2}
\end{equation}
where \teq{\psi (x)=d/dx [ \log_e\Gamma (x)]} is the Digamma function.  Setting
\teq{\mu =(\delta+1)/2}, since the upper limit to the \teq{P_{\delta}} 
can be replaced by infinity, one readily arrives at the result
\begin{eqnarray}
   \ndotssc (\erg )\Bigl\vert_{\rm NT} & = & 2^{(\delta -1)/2}
   \, \dover{(\delta-1)^2}{(\delta+1)}
   \Gamma \biggl( \dover{\delta}{4} + \dover{19}{12}\biggr)\,
   \Gamma \biggl( \dover{\delta}{4} - \dover{1}{12}\biggr)
   \;    \Biggl\{ \biggl[  \log_e\dover{\chi}{2} + \dover{2}{\delta +1}\nonumber\\[-5.5pt]
 \label{eq:SSCnt_highchi}\\[-5.5pt]
    && \hbox{\hskip 0pt} 
        - \dover{1}{2}  \psi \biggl( \dover{\delta}{4} + \dover{19}{12}\biggr)
        - \dover{1}{2} \psi \biggl( \dover{\delta}{4} - \dover{1}{12}\biggr)\, \biggr]
   \, G_{\delta}(1)+ H_{\delta}(1) \Biggr\}\;
   \dover{\Ndotssc}{\chi^{(\delta +1)/2}}\quad ,\quad \chi\,\gg\, 1\;\;  .\nonumber
\end{eqnarray}
The logarithmic term will clearly dominate for extremely large values of \teq{\chi}.


\section{The Synchrotron Self-Compton Rate for Thermal Electrons}
 \label{sec:appendixB}
 
The comparison of synchrotron self-Compton (SSC) models invoking shock
acceleration with burst spectra requires also the SSC rate for
relativistically hot thermal particles, which implies using a four-dimensional
integral defined by inserting the thermal contribution to
Eq.~(\ref{eq:elec_dist}) into Eq.~(\ref{eq:synch_emiss}) and then the
result subsequently into Eq.~(\ref{eq:IC_emiss}). Rather than
laboriously reducing the integral along the lines of the developments of
Appendix A, the result in Eq.~(\ref{eq:sscfinal_app}) can be used to
advantage.  Taking the limit \teq{\delta\to\infty} generates the SSC
emissivity for monoenergetic electrons with Lorentz factor
\teq{\gammin}.  The result can then be convolved with the thermal
electron distribution to yield the required emissivity.  As \teq{\delta}
becomes very large, \teq{P_{\delta}(z)} approaches zero at all positive
\teq{z} less than unity, so that the integration over
\teq{P_{\delta}(z)} in Eq.~(\ref{eq:sscfinal_app}) does not contribute. 
Using the result
\begin{equation}
    \lim_{\delta\to\infty} \delta Q_{\delta}(z)\;\equiv\; j(z)
    \; =\; 2(z+2)(z-1) - 2(1+2z)\,\log_e z \quad ,
 \label{eq:Qlimit}
 \end{equation}
the integration over the thermal electron distribution produces a double integral
that can be evaluated by reversing the order of integrations, yielding
\begin{equation}
   \ndotssc (\erg )\Bigl\vert_{\rm TH} \; = \; \Ndotsscth \int_{0}^{\infty} dy \, K_{5/3}(y)\, 
   {\cal J}\Bigl( \Bigl[\dover{\chit}{y}\Bigr]^{1/4}\Bigr)\quad ,
 \label{eq:sscfinal_th}
\end{equation}
where \teq{\Ndotsscth} is given in Eq.~(\ref{eq:Ndotsscdef}), 
\teq{\chit = \erg\, B_{\rm cr}/[6B_{\perp}\gammat^4]}, and
\begin{eqnarray}
   {\cal J}(\kappa ) &=& \dover{1}{\kappa}
   \int_1^{\infty} \dover{dt}{t^2} e^{-\kappa t}\, j(1/t^4)\nonumber\\[-5.5pt]
 \label{eq:calJdef}\\[-5.5pt]
   &\equiv & 2\dover{e^{-\kappa}}{\kappa} \int_0^{\infty} 
   \dover{e^{-v}s(v)}{\kappa +v}\, dv
     \; =\; 2\dover{e^{-\kappa}}{\kappa^3} \int_0^{\infty} 
   \dover{e^{-v} \, v^2\, s(v)}{\kappa +v}\, dv \quad ,\nonumber
\end{eqnarray}
where
\begin{equation}
   s(v)\; =\; \dover{v^9}{9!} + \dover{v^5}{5!}\, \Bigl( 8\psi (6) - 8 \log_ev +1 \Bigr)
   + 2 v \, \Bigl(2 \psi (2) - 2\log_ev -1 \Bigr)\quad .
 \label{eq:sdef}
\end{equation}
The first integral defining \teq{\cal J} can be expressed as a sum of
incomplete Gamma functions and one of their derivatives, using
identities 3.381.3 and 4.358.1 of Gradshteyn and Ryzhik (1980), though
this does not expedite the evaluation of \teq{\cal J}.  However, the
integral representation of the incomplete Gamma function in 8.353.3 of
Gradshteyn and Ryzhik (1980) proves useful, leading to the alternative
integral forms posited on the second line of Eq.~(\ref{eq:calJdef}).  
These latter forms are readily amenable to numerical computation in a
fashion similar to evaluation of Euler's integral form for the Gamma
function.  Note that the first form on the second line is expedient for
\teq{\kappa\lesssim 1} cases, while the second form on this line, which
was obtained using the fact that the integrals of \teq{v\, e^{-v} s(v)}
and \teq{v^2\, e^{-v} s(v)} on \teq{[0,\infty )} are identically zero,
is advantageous for evaluation when \teq{\kappa\gtrsim 1}.  Asymptotic
forms for \teq{\cal J} are simply obtained: \teq{{\cal J}(\kappa)\approx
1184/(225\kappa )} for \teq{\kappa\ll 1}, and \teq{{\cal
J}(\kappa)\approx 32 e^{-\kappa}/\kappa^4} for \teq{\kappa\gg 1}. The
appropriate asymptotic limits of Eq.~(\ref{eq:sscfinal_th}) can then be
quickly determined:
\begin{equation}
   \ndotssc (\erg )\Bigl\vert_{\rm TH} \; = \; \Ndotsscth
   \cases{ \dover{21}{25}\; 2^{2/3} \Bigl[ \Gamma (2/3)\Bigr]^2\;\chit^{-2/3}\;\; , &
       $\quad \chit\ll 1\quad , \vphantom{\Biggl(}$\cr
          \dover{8\pi}{\sqrt{5}}\; \dover{2^{9/10}}{\chit^{4/5}}\;
          \exp \biggl\{ -\dover{5}{2^{8/5}}\; \chit^{1/5} \biggr\}\;\; , &
        $\quad \chit\gg 1\quad , \vphantom{\Biggl(}$\cr}
 \label{eq:SSCTHlimits}
\end{equation}
where the low argument power-law dependence for the Bessel function
\teq{K_{5/3}(y)} is used when \teq{\chit\ll 1}, and the method of
steepest descents is employed to derive the \teq{\chit\gg 1} case, for
which the exponential portion of \teq{K_{5/3}(y)} is sampled.

Note that the SSC emissivity for a monoenergetic electron distribution
can be determined by omitting the convolution with the thermal
distribution, so that Eq.~(\ref{eq:Qlimit}) can be used efficaciously to
yield
\begin{equation}
   \ndotssc (\erg )\Bigl\vert_{\rm MONO} \; = \; \Ndotssc \int_{0}^{\infty} dy \, K_{5/3}(y)\, 
   j\Bigl( \dover{\chit}{y} \Bigr)\quad .
 \label{eq:sscfinal_mono}
\end{equation}
As \teq{j(z)} is a monotonically decreasing function of \teq{z} on
\teq{(0,1]}, for \teq{\chi\lesssim 1} it flattens the SSC spectrum
relative to that for synchrotron emission, and so effects a broadening
of the \teq{\nu F_{\nu}} peak.  This monoenergetic formula can be
convolved with arbitrary electron distributions to obtain resultant
synchrotron emissivities.  By the same token, Eq.~(\ref{eq:sscfinal_th})
can be used to facilitate a Laplace transform approach for special cases
for the electron distribution.


\clearpage

\end{document}